\documentclass[a4paper,fleqn]{cas-sc}

\usepackage[numbers]{natbib}
\usepackage{blkarray}%
\usepackage{booktabs}
\usepackage{float}

\def\tsc#1{\csdef{#1}{\textsc{\lowercase{#1}}\xspace}}
\tsc{WGM}
\tsc{QE}
\tsc{EP}
\tsc{PMS}
\tsc{BEC}
\tsc{DE}

\begin{document}
\let\WriteBookmarks\relax
\def\floatpagepagefraction{1}
\def\textpagefraction{.001}
\shorttitle{Ki-L mRNA Optimization}
\shortauthors{Zheng Gong et al.}

\title [mode = title]{mRNA Design and Optimization with Deep Knowledge‑Infused Approach}                   



\author[1]{Zheng Gong}
\fnmark[1]
\ead{hudenjear@gmail.com}
\credit{Conceptualization, Methodology, Validation, Data Curation, Writing - Original Draft, Visualization}

\affiliation[1]{organization={Tsinghua Shenzhen International Graduate School},
                city={Shenzhen},
                postcode={518055}, 
                state={Guangdong},
                country={China}}

\author[1]{Ziyi Jiang}
\fnmark[1]
\ead{jiang-zy20@mails.tsinghua.edu.cn}
\credit{Conceptualization, Validation, Writing - Review and Editing}

\author[1]{Weihao Gao}
\ead{gwh20@mails.tsinghua.edu.cn}
\credit{Conceptualization, Writing - Review and Editing}

\author[1]{Yuanyuan Wang}
\ead{wang-yy24@mails.tsinghua.edu.cn}
\credit{Validation}

\author[1]{Zhining Cai}
\ead{caizn24@mails.tsinghua.edu.cn}
\credit{Validation}

\author[1]{Zhuo Deng}
\ead{dengzhuo150@gmail.com}
\credit{Writing - Review and Editing}

\author[1,2]{Lan Ma}
\cormark[1]
\ead{malan@sz.tsinghua.edu.cn}
\credit{Funding acquisition, Supervision}

\affiliation[2]{organization={Shenzhen Bay Laboratory},
                city={Shenzhen},
                postcode={518107}, 
                state={Guangdong},
                country={China}}

\cortext[cor1]{Corresponding author}
\fntext[fn1]{These authors contributed equally to this work.}

\begin{abstract}
The mRNA optimization is essential for mRNA vaccines, therapies, and industrial protein production. Based on current explorations, an ideal optimization approach should simultaneously (i) prevent unintended amino-acid changes, (ii) optimize multiple, biologically relevant objectives, and (iii) retain computational efficiency. However, existing methods are forced to trade off between these perspectives, forming an "impossible triangle." We present RNop, a knowledge‑infused Transformer that integrates mechanism‑aligned losses to address this problem. By encoding biological prior knowledge in losses, RNop makes knowledge infusion explicit and controllable across optimization focus. Trained on over 6 million sequences, \textit{in silico} analyses show RNop resolves the "impossible triangle" of mRNA optimization with absolute sequence fidelity, significantly improved biological metrics, and high throughput. In \textit{in vitro} validation, it can deliver up to 2.28‑fold expression gain. Ablation studies reveal how each prior contributes to targeted improvements, yielding mechanism‑level interpretability. RNop represents a shift in mRNA optimization methodology: by infusing explicit and interpretable knowledge, the "black-box" mRNA design can be transformed into a predictable, explainable engineering problem. RNop is designed as an extensible platform: additional biological priors can be incorporated as modular, mechanism-aligned loss functions, enabling future development and adaptation to related sequence design problems. 
\end{abstract}




\maketitle
\begin{figure}[hbp]
  \centering
  \includegraphics[width=0.88\linewidth]{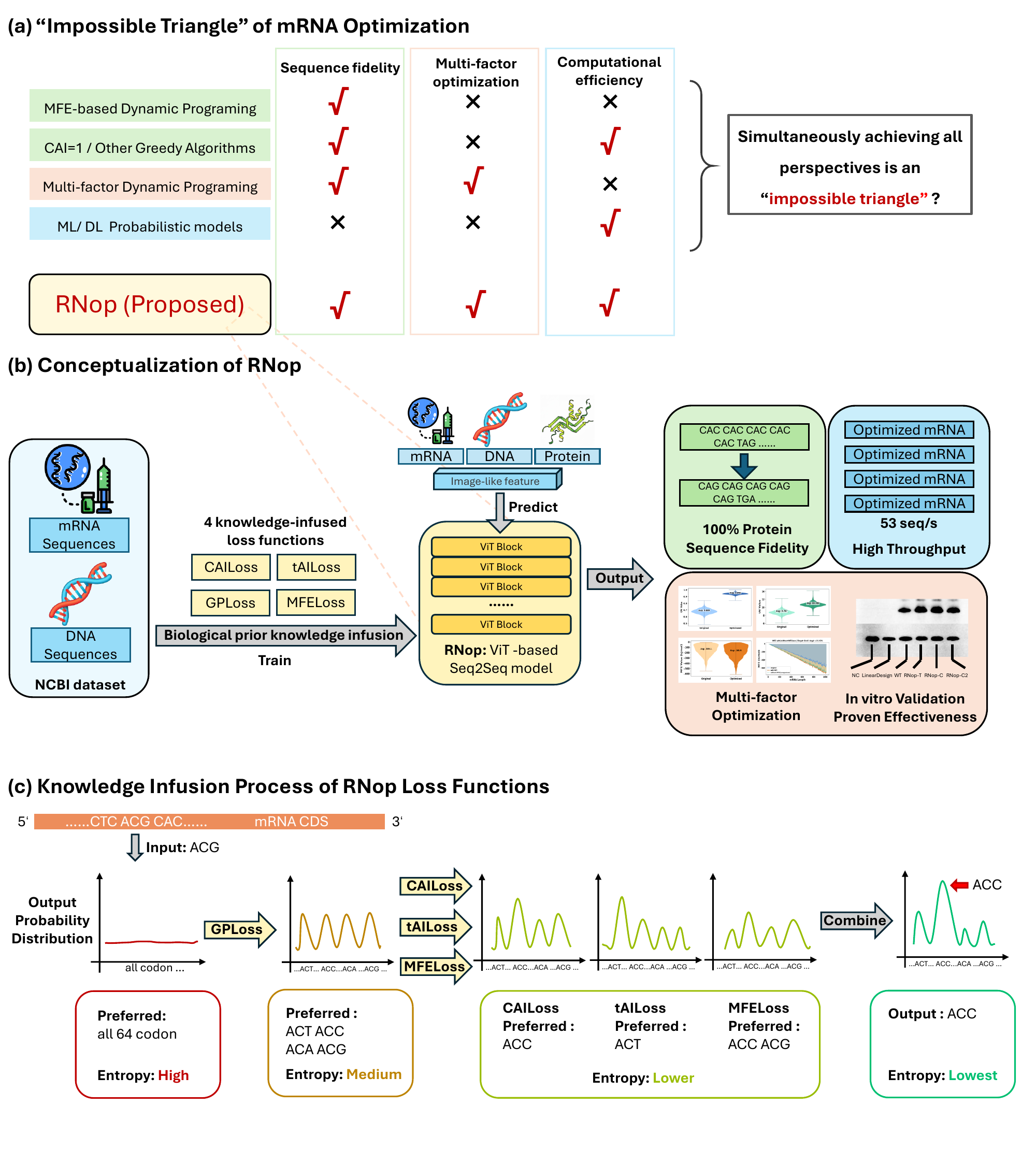}
  \caption{(a) \textbf{The "impossible triangle" of mRNA optimization.} Simultaneously achieving high sequence fidelity, computational efficiency, and a wide range of optimization objectives remains a significant challenge for current mRNA design methods. We propose a new knowledge-infused deep learning method to achieve all factors concurrently. (c) \textbf{The conceptualization of RNop}: Our methods include 4 knowledge-infused loss functions leveraging biological prior knowledge. RNop can process various sequences through a ViT-based Sequence-to-Sequence architecture. The optimized sequences exhibit great expression improvements in both computerized and biological experiments. (c) \textbf{Effect of infused Knowledge}. After the input process, each codon has a probability space spanning the whole vocabulary. The infused knowledge helps RNop models reduce the information entropy, providing the preferred codon for the input codon in specific species. The output is the optimal, synonymous codon.}
  \label{figmRNAflow}
\end{figure}

\section{Introduction}
The design and optimization of messenger RNA (mRNA) sequences represent a critical consideration in modern biological and biomedical research. As the mRNA sequence dictates translation efficiency and subsequent protein yield, challenges in achieving robust heterologous expression of target genes have driven the development of mRNA optimization methods. This optimization is now indispensable for a wide range of biotechnological goals, encompassing large-scale protein production, gene and cell therapies, synthetic biology circuits, and new medicine strategies. The performance of mRNA molecules as therapeutic agents or research tools hinges on their sequence characteristics, directly impacting efficacy and production scalability. 
Advancements in mRNA design and optimization also contribute to recombinant protein manufacturing and drug discovery pipelines \cite{Jain2023}. Given the expanding therapeutic and industrial applications of mRNA \cite{Paremskaia2024bn}, refining optimization techniques remains an area of intense investigation. 

Current mRNA optimization algorithms can be assessed based on several key criteria as shown in Fig.\ref{figmRNAflow} (a): (1) \textbf{fidelity} (preventing unintended amino acid changes), (2) \textbf{the scope of optimization variables considered} (optimality in multiple biological metrics), and (3) \textbf{computational efficiency} (speed and scalability). However, they are forced to trade off between these essential criteria, constrained by a central challenge, which we term the "impossible triangle" of mRNA optimization. To date, no existing method can simultaneously satisfy all three objectives, creating a significant bottleneck for advanced mRNA engineering. 

This "impossible triangle" arises from the inherent limitations of current design paradigms. Algorithmic methods like Dynamic Programming (DP) can ensure perfect sequence fidelity \cite{Terai2016, TANEDA20201811, Nikita2024mRNAid} and incorporate multiple optimization factors \cite{Zhang2023}. However, their computational complexity makes them slow for long sequences and incapable of high-throughput applications. Conversely, deep learning models \cite{Brixi2025.02.18.638918,Fu2020, Goulet2023} offer excellent computational efficiency and scalability. However, as probabilistic models, they struggle to guarantee absolute sequence fidelity, often introducing undesired protein mutations. Furthermore, typically, they learn optimization rules implicitly from natural sequence data \cite{Fu2020, Goulet2023,deepmRNAfullopt2023, doi:10.1126/science.adr8470}, which limits their ability to explicitly and concurrently optimize for multiple, well-studied biological metrics.

From a technical perspective, the central challenge is enforcing strict synonymous constraints within deep learning models while still supporting rich, multi-objective optimization. Conventional sequence-to-sequence or language-model-based approaches operate in codon token space and can only reduce, but not eliminate, protein-level mutations; post-hoc "repair" of mutated codons often destroys the jointly optimized distribution and offers no principled way to balance fidelity against exploration. In addition, when all biological factors are entangled in a single loss, it becomes difficult to attribute improvements to specific mechanisms or to tune their relative influence. These limitations motivate the need for a mechanism that can encode biological priors as differentiable constraints at the protein level and expose their effects in a controllable way.

The root of this problem lies in the methodology. Most deep learning approaches treat the optimization process as a black box \cite{doi:10.1126/science.adr8470}. These models learn latent knowledge from sequences, 
without leveraging accumulated biological knowledge. Therefore, we propose a different paradigm: we explicitly infuse established biological prior knowledge in the model training process through loss functions. This approach, knowledge-infused learning (K-iL), can leverage prior knowledge to enhance model performance and has achieved success in physics and other fields. \cite{raissi2019physics,pandya2022infusedheart,kursuncu2019knowledge} To the best of our knowledge, RNop is the first application of the K-iL in mRNA optimization, extending the spirit of physics-informed neural networks (PINNs) \cite{raissi2019physics} to mRNA sequence design.

To be more specific, we introduce RNop, a novel deep learning framework that embodies this knowledge-infused methodology to solve the ``impossible triangle''. At its core, RNop introduces a matrix-based constraint enforcement mechanism that enables strict synonymous CDS optimization in deep learning. This mechanism is instantiated on top of a Vision Transformer (ViT) backbone that treats mRNA optimization as an image-to-image translation problem. Meanwhile, RNop learns from explicit, mechanism-aligned loss functions rather than directly regressing target sequences. 
We developed four specialized loss functions: Gene-Protein Loss (GPLoss), CAILoss, tAILoss, and MFELoss. One of Our key technical contributions is the linear-algebraic formulation of these losses, which makes protein-gene correlation knowledge natively incorporated in deep learning. GPLoss operates directly on the codon probability space and penalizes non-synonymous probability mass, thereby enforcing protein-level fidelity during optimization. This is fundamentally different from post-processing ``repair'' strategies that replaces mutated codons after decoding \cite{Fu2020}, which may cancel optimization gains. Meanwhile, in RNop, the GPLoss weight provides such control: sufficiently high weights yield strict synonymy, while lower weights intentionally allow limited exploration when desired. Building on the same design philosophy, CAILoss and tAILoss incorporate codon-usage bias \cite{10.1093/nar/15.3.1281} and tRNA adaptation \cite{tai1,tai2} for a target species, and MFELoss incorporates secondary-structure considerations by minimizing predicted MFE, which is associated with mRNA stability and potentially longer half-lives \cite{MATHEWS1999911, david2004, BUHR2016341}. Together, these losses enable one-time multi-objective optimization of full CDS sequences.

By encoding biological priors as objective-aligned losses, RNop makes knowledge infusion explicit and enables a controllable, multi-objective optimization that was previously unattainable. The brief framework of our methods is shown in Fig.\ref{figmRNAflow} (b), and a detailed description is given in Section. Method. We outline the optimization process in Fig.\ref{figmRNAflow} (c), in which codon "ACG" is selected as an example to show how the knowledge learned from four loss functions influences the output probability distribution during the inference. 

Then, we trained and validated RNop models on a dataset of 6,000,000 mRNA sequences. Extensive \textit{in silico} analysis demonstrates that RNop consistently improves all target metrics across multiple species while maintaining perfect sequence fidelity. Crucially, in \textit{in vitro} biological validation experiments, RNop-optimized sequences achieved up to a 2.28-fold increase in protein expression compared to commercially available and highly optimized sequences, confirming the real-world efficacy of RNop. With a processing speed of up to 44 sequences per second, RNop also solves the efficiency problem, enabling high-throughput design. Through this work, we provide a high-fidelity, interpretable, and scalable solution that breaks the "impossible triangle." Beyond a single model, RNop is designed as a modular and extensible platform: new biological priors and species-specific objectives can be incorporated as plug-in mechanism-aligned loss functions, and new network structures can be applied. We hope this project can support future development and adaptation to related sequence design problems. Our code and models are available at {https://github.com/HudenJear/RPLoss}.  

\subsection{Materials and Methods}

\subsubsection{Dataset}
\label{sec_dataset}
The genome data is downloaded from the NCBI database. We downloaded the gene sequences of Eukaryota and Bacteria from the NCBI Genome Database\cite{NCBI_Genome}. The Eukaryota and Bacteria sequences are filtered: only the complete and known mRNA (transcriptions) are included. The virus genome data is downloaded from the NCBI Virus Database\cite{NCBI_Virus}. The virus sequences are filtered: only the sequences that are complete and have lengths from 1 to 30000 were downloaded. All sequences were downloaded prior to May 1th 2025. In the initial stage, 21,194,000 sequences are downloaded. 

We preprocess these raw sequence data into the dataset. In this step, the sequences are cropped to the UTRs and CDS sequences. The duplicated CDS sequences (mRNA) are deleted to avoid data leaking for validation and testing. For the basic RNop model, the sequences over 45 and under 3072 nt are included. For the extended RNop model with doubled and 4-fold maximum lengths, the sequences over 45 nt and under 6144/12288 nt are included. 

To balance the data sources, we randomly sampled 2,000,000 sequences from each category: Eukaryota, Bacteria, and Virus. Therefore, we collected an mRNA sequence dataset containing 6,000,000 mRNA sequences for training, validation, and testing with balanced categories. Compared to 38000\cite{Goulet2023}, 42266\cite{Jain2023}, or 71820\cite{Nicole2023PredictionCodonBERT}, the amount of mRNA sequences is larger than previous codon optimization methods.

The sampled sequences are further divided into train, validation, and test sets, with partitions of 90\%, 1\%, and 9\%, respectively. There are 5,400,000 sequences in the training data, and 60,000 and 540,000 in the validation and test data. Meanwhile, to test the homologous optimization ability, another species-specified test set of 100,000 sequences (25,000 in virus, E.coli, mouse, and human) was built, using the data not included in the sampling step.

\subsubsection{Biological Experiment Setting}

In vitro experiments. The HEK293T cell line used in this study was preserved in our laboratory. 

\textbf{mRNA transfection:} HEK293T cells were cultured in DMEM medium(GIBCO, \#11965092) supplemented with 10\% fetal bovine serum (GIBCO, \#A5670701) and 1\% penicillin-streptomycin(YEASEN ,\#60162ES ), and maintained in a 37°C incubator with 5\%  carbon dioxide. mRNA was transfected into HEK293T cells using Lipofectamine™ MessengerMAX™ Transfection reagent(Invitrogen, \#LMRNA015) following the manufacturer's instructions. At the appropriate time after transfection, cells and cell lysates were collected for fluorescence and protein detection.

\textbf{In Vitro Protein Detection:} To detect the in vitro expression of fluorescent proteins EGFP and ECFP, HEK293T cells were seeded in 24-well plates, with 1 $\mu g$ of mRNA transfected into each well. At 24 h and 48 h post-transfection, the fluorescence intensity of cells in each well was measured using a fluorescence spectrophotometer. Meanwhile, cells were harvested at the same time points, lysed, and subjected to western blot analysis with EGFP(ABCAM, \#ab184601) and ECFP(ABBKINE, \#abt2280) antibodies. Additionally, western blot detection was performed on the optimized sequences of other target proteins using His antibodies(PROTEINTECH, \#66005-1-Ig).

\subsubsection{Brief look of RNop}

Our RNop includes a coding method for various sequences, 4 knowledge-infused loss functions incorporating biological prior knowledge, and a new Transformer-based deep learning model, the RNop model. 

The coding method is the cornerstone of our method. With this coding rule, we are able to encode multiple types of gene sequences into uniformly formatted feature maps. Then, to leverage the prior knowledge from biology, we create four loss functions to guide the deep learning model training: GPLoss, CAILoss, tAILoss, and MFELoss. 
Furthermore, the RNop model is designed to optimize the mRNA CDS with multiple optimization factors, high sequence fidelity, and high throughput. Through our coding method, the RNop model obtains the ability to process mRNA, DNA CDS, or protein sequences and optimize gene sequences species-specifically.

\subsubsection{mRNA Optimization Model: RNop}

The mRNA sequences are simpler on the one hand-- they usually have less vocabulary--and complicated on the other hand--they are more strict with length and coding. Therefore, we designed a new model for mRNA optimization: the RNop model. The RNop model is a sequence-to-sequence model that consists of the Transformer and convolutional network layers and leverages the Vision Transformer (ViT) \cite{dosovitskiy2021,liu2021swin} architecture. The choice of the ViT model is not an arbitrary decision. We have tested a wide range of deep learning models and found that the ViT model is the only model that can fully avoid errors and utilize the feature of knowledge‑infused loss functions.

\begin{figure}
    \centering
    \includegraphics[width=0.95\linewidth]{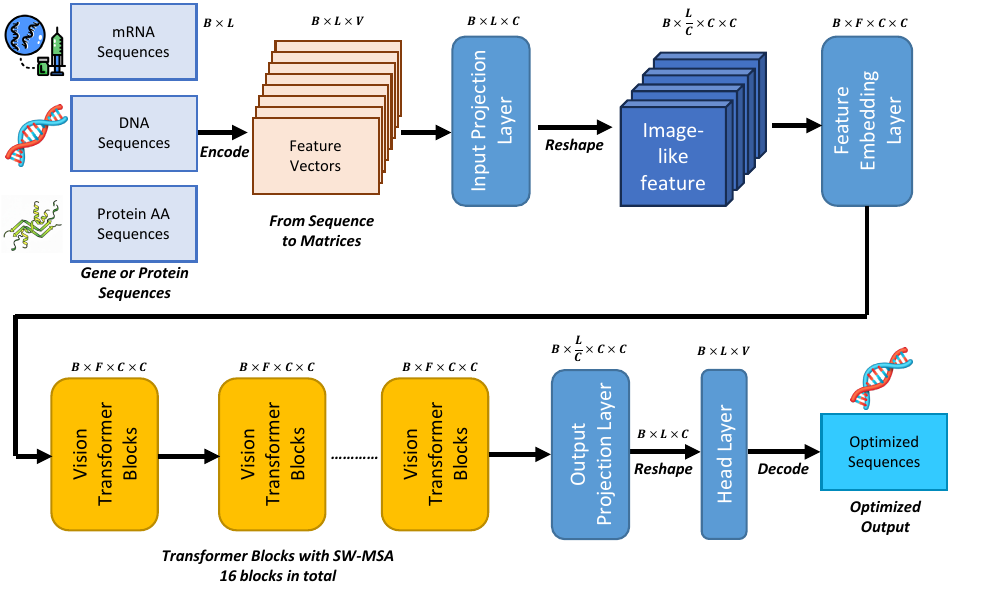}
    \caption{\textbf{The architecture of our RNop model}: (1) RNop model uses our coding method to encode multiple kinds of gene sequences to feature matrices. (2) The feature matrices are processed and reshaped to image-like features. (3) Then, the model can process these features using Vision Transformer structures. (4) The features are unpacked and restored to sequences and decoded to the desired optimal sequences.}
    \label{figarch}
\end{figure}

The RNop model leverages the ViT architecture to accomplish the sequence-to-sequence task. Its architecture is shown in Fig.\ref{figarch}. As discussed above, we can choose mRNA, DNA, or amino acids coding for the input. After the encoding process, the $<EOS>$ and $<PAD>$ are added to the input, and the sequences are transformed to codon possibility space matrices. We assume the input sequences $I$ have the shape of $B\times L\times V$, where the $B$, $L$, and $V$ represent the batch size, sequence length, and vocabulary size. To standardize the input dimension, the input $I$ is processed by an input projection layer to adjust its shape to $ X_{1} \in \mathbb{R}^{B\times L\times C}$, where $C$ represents the projection channel. The feature $X_1$ is reshaped to $X_2 \in \mathbb{R}^{B\times \frac{L}{C}\times C\times C}$, which is akin to an image with $\frac{L}{C}$ channels, height $C$ and width $C$.

Subsequently, the image-like feature is processed by a feature embedding layer, which is a convolutional layer with a stride of 1 and a kernel size of 3. (We observed severe information loss when we applied the non-overlapping patch embedding layer.) The processed feature is $X_3 \in \mathbb{R}^{B\times F\times C\times C}$, where $F$ is the feature embedding dimension. Then, $X_3$ is processed by 16 Shifted Window-based Multihead Self-Attention layers\cite{liu2021swin}. This attention mechanism ensures that the global feature and local feature can be extracted and fused. The feature size is kept the same during this process, and no downsampling or upsampling is applied. We call this component the ViT encoder. The output of the last ViT encoder layer is $O_1\in \mathbb{R}^{B\times F\times C\times C}$. Then, the $O_1$ is processed by a point-wise convolutional layer, modifying its dimension from $F$ to $\frac{L}{C}$ and further reshaped back to $O_2 \in  \mathbb{R}^{B\times L\times C}$.

Consequently, we utilize a head layer to adjust the last dimension of the $O_2$ from $C$ to $V$ to give the final output $O\in \mathbb{R}^{B\times L\times V}$. Then the output can be decoded to mRNA sequences or used directly in the loss function calculation for gradient descent.

\subsubsection{Data Representation: Coding Rules}
\label{ssec:cr}

We designed a coding rule to standardize mRNA and other sequences for neural networks. While some LLMs use tokenizers and K-mers\cite{Muhammad2020}, they can be redundant due to the limited number of codons. 
Therefore, we propose a new coding method that directly converts sequences into numerical matrices to avoid redundancy.

Generally, the mRNA sequences consist of four ribonucleotides: A, U, C, and G. Meanwhile, most RNA datasets (including NCBI\cite{NCBI_Genome} datasets) use "T" to replace "U" to keep the representation the same as the DNA sequences. Thus, following this conventional usage, we represent RNA sequences accordingly to include both mRNA and DNA CDS. With a fixed order "$TCAG$" and 3 nucleotides in one codon, we can settle the position of each codon in the matrices. This enables us to map one codon to its coding protein using linear algebra by matrix multiplication. The vocabulary of this coding rule is $4^3=64$, which is also the width of the transformed matrices. Since the length is the sequence length, the matrices can represent the probability space of input and output sequences. The detailed mapping method of this coding rule can be referred to in Supplementary information.

Additionally, this coding rule is compatible with uncertain codons that are denoted as “N” or “RYMKSWHBVDN.” With five alphabets representing RNA, the vocabulary size is $5^3=125$. Although using more alphabets (e.g., RYMKSWHBVDN) provides richer information than a singleton “N,” it would inflate the vocabulary size to $15^{3}=3375$. As they are really rare in the sequences, we decided to only include the "N" codons in the RNop model.

Furthermore, RNA sequence lengths vary a lot. Alongside codon symbols, we introduce an $<EOS>$ (End Of Sequence) symbol at the end of RNA sequences. To ensure consistent input and output lengths for neural networks, we also add $<PAD> $ (Padding) coding symbols. Working with GPloss, these tokens can avoid length change in output sequences. 

\subsubsection{GPLoss: Avoiding Mutation Through Coding Rule}
\label{ssec:rp}

GPLoss is built on the coding rule and is designed to enforce \textit{sequence fidelity} at the protein level via a differentiable, matrix-based constraint. Using the codon-to-amino-acid mapping, GPLoss penalizes output probability mass assigned to codons that would translate into an amino acid different from that of the input codon at the same position. Unlike post-hoc ``repair'' approaches that modify decoded sequences after generation, this formulation constrains the probability space during training and optimization inference and thus preserves the joint, multi-objective behavior of the model. With a sufficiently large loss weight, this mechanism strongly suppresses non-synonymous mutations and preserves the translated protein sequence. Conversely, reducing the GPLoss weight relaxes this constraint and can allow a controlled amount of amino-acid changes. In practice, tuning the GPLoss weight provides a direct handle on the mutation rate, ranging from 0\% strictly to higher values when intentionally permitted.

\textbf{Worked example (standard nucleotides only)}

Assume three input codons: UGC, CAU, and AAG. These codons encode the amino acids Cysteine, Histidine, and Lysine, respectively. To illustrate the GPLoss computation end-to-end, we include the synonymous codons for these amino acids and additionally Phenylalanine as an extra example and construct the input codon matrix $I$ as follows.
\begin{footnotesize}
    \begin{equation}
I=\begin{blockarray}{ccccccccc}
&UGU&UGC&CAU&CAC&AAA&AAG&UUU&UUC\\
\begin{block}{c[cccccccc]}
input1: AAG & 0 & 0 & 0 & 0 & 0 & 1 & 0 & 0\\
input2: UGC & 0 & 1 & 0 & 0 & 0 & 0 & 0 & 0\\
input3: CAU & 0 & 0 & 1 & 0 & 0 & 0 & 0 & 0\\
\end{block}
\end{blockarray}
\end{equation}
\end{footnotesize}

Next, $I$ is multiplied by an amino-acid encoding matrix (denoted as $C$) that represents the mapping between codons and amino acids. For clarity in the computation, we include one not coded amino acid, Phenylalanine, in $C$ as well.
\begin{footnotesize}
    \begin{equation}
C=\begin{blockarray}{ccccccccc}
&UGU&UGC&CAU&CAC&AAA&AAG&UUU&UUC\\
\begin{block}{c[cccccccc]}
Cys & 1 & 1 & 0 & 0 & 0 & 0 & 0 & 0\\
His & 0 & 0 & 1 & 1 & 0 & 0 & 0 & 0\\
Lys & 0 & 0 & 0 & 0 & 1 & 1 & 0 & 0\\
Phe & 0 & 0 & 0 & 0 & 0 & 0 & 1 & 1\\
\end{block}
\end{blockarray}
\end{equation}
\end{footnotesize}

Multiplying the input matrix by the transpose of the encoding matrix yields the amino-acid identity for each input codon. We denote this intermediate matrix as $E$:
\begin{footnotesize}
   \begin{equation}
    E=\begin{blockarray}{cccccccc}
    \begin{block}{[cccccccc]}
     0 & 0 & 0 & 0 & 0 & 1 & 0 & 0\\
     0 & 1 & 0 & 0 & 0 & 0 & 0 & 0\\
     0 & 0 & 1 & 0 & 0 & 0 & 0 & 0\\
    \end{block}
    \end{blockarray}
    \cdot
    \begin{blockarray}{cccc}
    Cys & His & Lys & Phe \\
    \begin{block}{[cccc]}
     1 & 0 & 0 & 0 \\
     1 & 0 & 0 & 0 \\
     0 & 1 & 0 & 0 \\
     0 & 1 & 0 & 0 \\
     0 & 0 & 1 & 0 \\
     0 & 0 & 1 & 0 \\
     0 & 0 & 0 & 1 \\
     0 & 0 & 0 & 1 \\
    \end{block}
    \end{blockarray}
    =
    \begin{blockarray}{ccccc}
     & Cys & His & Lys & Phe \\
    \begin{block}{c[cccc]}
    input1: AAG & 0 & 0 & 1 & 0 \\
    input2: UGC & 1 & 0 & 0 & 0 \\
    input3: CAU & 0 & 1 & 0 & 0 \\
    \end{block}
    \end{blockarray}
\end{equation} 
\end{footnotesize}

We then multiply $E$ by $C$ to obtain a mask over codons that are synonymous with the amino acid encoded by each input position. The resulting matrix is denoted as $P$.

\begin{footnotesize}
    \begin{equation}
    P=\begin{blockarray}{cccc}
     Cys & His & Lys & Phe \\
    \begin{block}{[cccc]}
     0 & 0 & 1 & 0 \\
     1 & 0 & 0 & 0 \\
     0 & 1 & 0 & 0 \\
    \end{block}
    \end{blockarray}
    \cdot
    \begin{blockarray}{cccccccc}
    \begin{block}{[cccccccc]}
     1 & 1 & 0 & 0 & 0 & 0 & 0 & 0\\
     0 & 0 & 1 & 1 & 0 & 0 & 0 & 0\\
     0 & 0 & 0 & 0 & 1 & 1 & 0 & 0\\
     0 & 0 & 0 & 0 & 0 & 0 & 1 & 1\\
    \end{block}
    \end{blockarray}
    =
    \begin{blockarray}{cccccccc}
    \begin{block}{[cccccccc]}
     0 & 0 & 0 & 0 & 1 & 1 & 0 & 0\\
     1 & 1 & 0 & 0 & 0 & 0 & 0 & 0\\
     0 & 0 & 1 & 1 & 0 & 0 & 0 & 0\\
    \end{block}
    \end{blockarray}
\end{equation}
\end{footnotesize}

After the RNop forward pass, the model outputs a probability distribution over codons at each position. For illustration, we show the sub-matrix corresponding to the codons above and denote it as $O$.
\begin{footnotesize}
    \begin{equation}
O=\begin{blockarray}{ccccccccc}
&UGU&UGC&CAU&CAC&AAA&AAG&UUU&UUC\\
\begin{block}{c[cccccccc]}
Output1 & 0 & 0.1 & 0.1 & 0.1 & 0.3 & 0.4 & 0 & 0\\
Output2 & 0 & 0.1 & 0.1 & 0.1 & 0.3 & 0.4 & 0 & 0\\
Output3 & 0 & 0.1 & 0.1 & 0.1 & 0.3 & 0.4 & 0 & 0\\
\end{block}
\end{blockarray}
\end{equation}
\end{footnotesize}

We take the element-wise product between $O$ and the synonymous-codon mask $P$ to obtain the matrix $OP$:
\begin{footnotesize}
    \begin{equation}
\begin{aligned}
    OP&=\begin{blockarray}{cccccccc}
UGU&UGC&CAU&CAC&AAA&AAG&UUU&UUC\\
\begin{block}{[cccccccc]}
 O_{11}*P_{11} & O_{12}*P_{12} & O_{13}*P_{13} & O_{14}*P_{14} & O_{15}*P_{15} & O_{16}*P_{16} &O_{17}*P_{17} & O_{18}*P_{18}\\
 O_{21}*P_{21} & O_{22}*P_{22} & O_{23}*P_{23} & O_{24}*P_{24} & O_{25}*P_{25} & O_{26}*P_{26} &O_{27}*P_{27} & O_{28}*P_{28}\\
 O_{31}*P_{31} & O_{32}*P_{32} & O_{33}*P_{33} & O_{34}*P_{34} & O_{35}*P_{35} & O_{36}*P_{36} &O_{37}*P_{37} & O_{38}*P_{38}\\
\end{block}
\end{blockarray}
\\
&=\begin{blockarray}{cccccccc}
\begin{block}{[cccccccc]}
 0 & 0 & 0 & 0 & 0.3 & 0.4 & 0 & 0\\
 0 & 0.1 & 0 & 0 & 0 & 0 & 0 & 0\\
0 & 0 & 0.1 & 0.1 & 0 & 0 & 0 & 0\\
\end{block}
\end{blockarray}
\end{aligned}
\end{equation}
\end{footnotesize}

Summing $OP$ over the codon dimension yields a vector $L$, which measures how much probability mass the model assigns to synonymous codons for each input position.

\begin{equation}
    L=\begin{blockarray}{c}
\begin{block}{[c]}
 \sum^{N}_j(O_{1j}*P_{1j}) \\
 \sum^{N}_j(O_{2j}*P_{2j})\\
 \sum^{N}_j(O_{3j}*P_{3j})\\
\end{block}
\end{blockarray}
=\begin{blockarray}{c}
\begin{block}{[c]}
 0.7 \\
 0.1\\
 0.2\\
\end{block}
\end{blockarray}
\end{equation}
Because the output probabilities are normalized, GPLoss can be computed from $L$ as:

\begin{equation}
    GPLoss=\sum^N_i(1-L_i)
\end{equation}

\textbf{Ambiguous nucleotides ($N$)}

Some sequences contain ambiguous nucleotides denoted as $N$, indicating that a base is unknown (typically one of $\{T,C,A,G\}$ under our convention). When $N$ appears within a codon, the corresponding codon token becomes ambiguous and can represent multiple concrete codons.


Introducing $N$ expands the codon vocabulary from $4^3$ to $5^3$, which increases ambiguity in the loss computation. Below we describe two practical strategies for handling codons that contain $N$.

To make the discussion concrete, we add one additional input containing $N$ and consider the extreme case $NNN$. The input matrix $I$ becomes:
\begin{footnotesize}
    \begin{equation}
      I=\begin{blockarray}{cccccccccc}
&UGU&UGC&CAU&CAC&AAA&AAG&UUU&UUC&NNN\\
\begin{block}{c[ccccccccc]}
input1: AAG & 0 & 0 & 0 & 0 & 0 & 1 & 0 & 0 & 0\\
input2: UGC & 0 & 1 & 0 & 0 & 0 & 0 & 0 & 0 & 0\\
input3: CAU & 0 & 0 & 1 & 0 & 0 & 0 & 0 & 0 & 0\\
input4: NNN & 0 & 0 & 0 & 0 & 0 & 0 & 0 & 0 & 1\\
\end{block}
\end{blockarray}
\end{equation}
\end{footnotesize}

Since $NNN$ may correspond to any codon (and thus any amino acid), we expand the encoding matrix $C$ accordingly:
\begin{footnotesize}
    \begin{equation}
C=\begin{blockarray}{cccccccccc}
&UGU&UGC&CAU&CAC&AAA&AAG&UUU&UUC&NNN\\
\begin{block}{c[ccccccccc]}
Cys & 1 & 1 & 0 & 0 & 0 & 0 & 0 & 0 & 1\\
His & 0 & 0 & 1 & 1 & 0 & 0 & 0 & 0 & 1\\
Lys & 0 & 0 & 0 & 0 & 1 & 1 & 0 & 0 & 1\\
Phe & 0 & 0 & 0 & 0 & 0 & 0 & 1 & 1 & 1\\
\end{block}
\end{blockarray}
\end{equation}
\end{footnotesize}

After multiplication, the matrix $E$ contains the implied amino-acid identities:
\begin{footnotesize}
    \begin{equation}
E=\begin{blockarray}{ccccc}
     & Cys & His & Lys & Phe \\
    \begin{block}{c[cccc]}
    input1: AAG & 0 & 0 & 1 & 0 \\
    input2: UGC & 1 & 0 & 0 & 0 \\
    input3: CAU & 0 & 1 & 0 & 0 \\
    input4: NNN & 1 & 1 & 1 & 1 \\
    \end{block}
    \end{blockarray}
\end{equation}
\end{footnotesize}

\textbf{Solution A (allow $N$ in predictions).} In this option, we allow $N$-containing codons to appear in the model output and do not explicitly penalize them. The same workflow as the non-$N$ case is applied, and $E$ is multiplied by $C$ to obtain $P_A$:
\begin{footnotesize}
    \begin{equation}
    \begin{blockarray}{cccc}
     Cys & His & Lys & Phe \\
    \begin{block}{[cccc]}
     0 & 0 & 1 & 0 \\
     1 & 0 & 0 & 0 \\
     0 & 1 & 0 & 0 \\
     1 & 1 & 1 & 1 \\
    \end{block}
    \end{blockarray}
    \cdot
    \begin{blockarray}{ccccccccc}
    \begin{block}{[ccccccccc]}
     1 & 1 & 0 & 0 & 0 & 0 & 0 & 0 & 1\\
     0 & 0 & 1 & 1 & 0 & 0 & 0 & 0 & 1\\
     0 & 0 & 0 & 0 & 1 & 1 & 0 & 0 & 1\\
     0 & 0 & 0 & 0 & 0 & 0 & 1 & 1 & 1\\
    \end{block}
    \end{blockarray}
    =
    \begin{blockarray}{ccccccccc}
    \begin{block}{[ccccccccc]}
     0 & 0 & 0 & 0 & 1 & 1 & 0 & 0 & 1\\
     1 & 1 & 0 & 0 & 0 & 0 & 0 & 0 & 1\\
     0 & 0 & 1 & 1 & 0 & 0 & 0 & 0 & 1\\
     1 & 1 & 1 & 1 & 1 & 1 & 1 & 1 & 4\\
    \end{block}
    \end{blockarray}
\end{equation}
\end{footnotesize}

Note that the value 4 in the last row is incompatible with probability-like masking and would distort the loss. Therefore, before applying $P_A$ to the model output, we clamp the values to $[0,1]$ (via $torch.clamp()$). Assume $O^{\prime}$ is the corresponding model output:
\begin{footnotesize}
    \begin{equation}
      O^{\prime}=\begin{blockarray}{cccccccccc}
&UGU&UGC&CAU&CAC&AAA&AAG&UUU&UUC&NNN\\
\begin{block}{c[ccccccccc]}
Output1 & 0 & 0.1 & 0.1 & 0.1 & 0.3 & 0.2 & 0 & 0 & 0.2\\
Output2 & 0 & 0.1 & 0.1 & 0.1 & 0.3 & 0.2 & 0 & 0 & 0.2\\
Output3 & 0 & 0.1 & 0.1 & 0.1 & 0.3 & 0.2 & 0 & 0 & 0.2\\
Output4 & 0 & 0.1 & 0.1 & 0.1 & 0.3 & 0.2 & 0 & 0 & 0.2\\
\end{block}
\end{blockarray}
\end{equation}
\end{footnotesize}

Then $OP$, $L$, and $GPLoss$ are computed as follows:
\begin{footnotesize}
    \begin{equation}
\begin{aligned}
    OP=
\begin{blockarray}{ccccccccc}
\begin{block}{[ccccccccc]}
 0 & 0 & 0 & 0 & 0.3 & 0.2 & 0 & 0 & 0.2\\
 0 & 0.1 & 0 & 0 & 0 & 0 & 0 & 0 & 0.2\\
 0 & 0 & 0.1 & 0.1 & 0 & 0 & 0 & 0 & 0.2\\
 0 & 0.1 & 0.1 & 0.1 & 0.3 & 0.2 & 0 & 0 & 0.2\\
\end{block}
\end{blockarray}
    ,  L=\begin{blockarray}{c}
\begin{block}{[c]}
 0.7 \\
 0.3\\
 0.4\\
 1.0\\
\end{block}
\end{blockarray}
,GPLoss=\sum^N_i(1-L_i)
\end{aligned}
\end{equation}
\end{footnotesize}

This formulation reveals a drawback: if $N$ is allowed in the prediction space, the $NNN$ token can weaken or remove useful gradients, and it may also encourage the model to output $N$-containing codons even when the input codon is unambiguous. When this behavior is undesirable, we can either rely on additional loss terms or adopt Solution B below to discourage ambiguous predictions.

\textbf{Solution B (discourage $N$ in predictions).} Here we still accept $N$ in the \textit{input}, but we discourage $N$ in the \textit{output}. To do so, we modify the second-step encoding matrix by removing the amino-acid associations of $N$-containing codons, yielding a new matrix $C_B$:
\begin{footnotesize}
    \begin{equation}
P_B=
    \begin{blockarray}{cccc}
     Cys & His & Lys & Phe \\
    \begin{block}{[cccc]}
     0 & 0 & 1 & 0 \\
     1 & 0 & 0 & 0 \\
     0 & 1 & 0 & 0 \\
     1 & 1 & 1 & 1 \\
    \end{block}
    \end{blockarray}
    \cdot
    \begin{blockarray}{ccccccccc}
    \begin{block}{[ccccccccc]}
     1 & 1 & 0 & 0 & 0 & 0 & 0 & 0 & 0\\
     0 & 0 & 1 & 1 & 0 & 0 & 0 & 0 & 0\\
     0 & 0 & 0 & 0 & 1 & 1 & 0 & 0 & 0\\
     0 & 0 & 0 & 0 & 0 & 0 & 1 & 1 & 0\\
    \end{block}
    \end{blockarray}
    =
    \begin{blockarray}{cccccccccc}
    \begin{block}{[cccccccccc]}
     0 & 0 & 0 & 0 & 1 & 1 & 0 & 0& 0\\
     1 & 1 & 0 & 0 & 0 & 0 & 0 & 0& 0\\
     0 & 0 & 1 & 1 & 0 & 0 & 0 & 0& 0\\
     1 & 1 & 1 & 1 & 1 & 1 & 1 & 1& 0\\
    \end{block}
    \end{blockarray}
\end{equation}
\end{footnotesize}

With this modification, $NNN$ no longer trivially matches every amino acid, and the model receives a meaningful gradient signal when it assigns probability mass to $NNN$. Using the same output $O^{\prime}$, we obtain:

\begin{footnotesize}
    \begin{equation}
\begin{aligned}
    OP=
\begin{blockarray}{ccccccccc}
\begin{block}{[ccccccccc]}
 0 & 0 & 0 & 0 & 0.3 & 0.2 & 0 & 0 & 0\\
 0 & 0.1 & 0 & 0 & 0 & 0 & 0 & 0 & 0\\
 0 & 0 & 0.1 & 0.1 & 0 & 0 & 0 & 0 & 0\\
 0 & 0.1 & 0.1 & 0.1 & 0.3 & 0.2 & 0 & 0 & 0\\
\end{block}
\end{blockarray}
    ,  L=\begin{blockarray}{c}
\begin{block}{[c]}
 0.5 \\
 0.1\\
 0.2\\
 0.8\\
\end{block}
\end{blockarray}
,GPLoss=\sum^N_i(1-L_i)
\end{aligned}
\end{equation}
\end{footnotesize}

Solution B therefore encourages the model to avoid predicting codons containing $N$, leading to less ambiguous outputs. Nevertheless, we note that codons with $N$ in the original input may be problematic for downstream biological experiments and should be treated with caution.\\

\textbf{Formulation of GPLoss}

Because each position corresponds to a normalized codon distribution, the per-position GPLoss score is bounded by 1. We summarize the computation in Eq.~\eqref{eq:gploss1}. Here, $\textbf{O,P}$ denotes the model output and the synonymous-codon mask (as above), and $\textbf{P}$ denotes the encoding matrix. $L_{GP,i}$ is the per-position synonymous-mass score, and $GPLoss$ is the total loss. In our experiments, the GPLoss weight is set to $10^{-1}$ to achieve a zero mutation rate.

\begin{equation}
  \textbf{OP}=sum(\textbf{O} \cdot (\textbf{P}))=\begin{blockarray}{c}
\begin{block}{[c]}
L_{GP,1}\\
L_{GP,2}\\
...\\
L_{GP,n}\\
\end{block}
\end{blockarray}
, GPLoss=\sum^n_{i=1}(1-L_{GP,i})
\label{eq:gploss1}
\end{equation}

\subsubsection{CAILoss: Improving CAI by Penalizing Low CAI Codons}
\label{ssec:cai}
 
We leverage the CAI \cite{10.1093/nar/15.3.1281} to design CAILoss, which explicitly injects codon-usage-bias knowledge into RNop. Intuitively, CAILoss evaluates the CAI scores of the model's output distribution at each codon position and penalizes probability mass assigned to low-CAI codons, thereby encouraging the selection of codons preferred by the target species. 

The CAILoss parameters are initialized from species-specific codon-usage tables. We downloaded the raw data from the open-source GenScript database \cite{genscript_codon_table}. Currently, we have incorporated parameters for more than 10 species, and additional species can be added by including their codon-usage-bias profiles.

Following the same workflow as GPLoss, we replace the binary entries in the encoding matrix with codon-specific CAI values, yielding a CAI-weighted matrix $C_{CAI-raw}$:

\begin{footnotesize}
    \begin{equation}
      C_{CAI-raw}=
\begin{blockarray}{cccccccccc}
&UGU&UGC&CAU&CAC&AAA&AAG&UUU&UUC&NNN\\
\begin{block}{c[ccccccccc]}
Cys & 0.46 & 0.54 & 0 & 0 & 0 & 0 & 0 & 0 & 0\\
His & 0 & 0 & 0.57 & 0.43 & 0 & 0 & 0 & 0 & 0\\
Lys & 0 & 0 & 0 & 0 & 0.74 & 0.26 & 0 & 0 & 0\\
Phe & 0 & 0 & 0 & 0 & 0 & 0 & 0.58 & 0.42 & 0\\
\end{block}
\end{blockarray}
\end{equation}
\end{footnotesize}

Using raw CAI values directly would not ensure that the optimal choice yields a per-position score of 1. Therefore, we normalize the CAI values within each amino-acid group by dividing by the maximum CAI among synonymous codons to form a matrix $C_{CAI}$:

\begin{footnotesize}
    \begin{equation}
      C_{CAI}=
\begin{blockarray}{cccccccccc}
&UGU&UGC&CAU&CAC&AAA&AAG&UUU&UUC&NNN\\
\begin{block}{c[ccccccccc]}
Cys & \frac{0.46}{0.54} & 1 & 0 & 0 & 0 & 0 & 0 & 0 & 0\\
His & 0 & 0 & 1 & \frac{0.43}{0.57} & 0 & 0 & 0 & 0 & 0\\
Lys & 0 & 0 & 0 & 0 & 1 & \frac{0.26}{0.74} & 0 & 0 & 0\\
Phe & 0 & 0 & 0 & 0 & 0 & 0 & 1 & \frac{0.42}{0.58} & 0\\
\end{block}
\end{blockarray}
\end{equation}
\end{footnotesize}

We then apply $C_{CAI}$ within the same two-step workflow as GPLoss: the input is first processed by the original matrix $C$ to obtain the amino-acid identity, and $C_{CAI}$ is applied in the second multiplication step to weight synonymous codons by their normalized CAI values. Finally, CAILoss follows the same aggregation as GPLoss to produce the scalar loss. The computation is summarized in Eq.~\eqref{eq:cailoss1}.

\begin{equation}
  L=sum(\textbf{O} \cdot (\textbf{I} \cdot \textbf{C}^{T} \cdot \textbf{C}_{CAI})) =\begin{blockarray}{c}
\begin{block}{[c]}
L_{CAI,1}\\
L_{CAI,2}\\
...\\
L_{CAI,n}\\
\end{block}
\end{blockarray}
, CAILoss=\sum^n_{i=1}(1-L_{CAI,i})
\label{eq:cailoss1}
\end{equation}

\subsubsection{tAILoss: Improving tAI by Penalizing Low tAI Codons}
\label{ssec:tai}
The tAILoss targets translation efficiency by incorporating tRNA availability information across species. Following the tAI formulation in prior work \cite{tai1,tai2}, codons whose corresponding anticodons have higher tRNA copy numbers are treated as more preferred. We adapt the original tAI computation to match our coding rule by re-ordering codons to align with our input/output matrix convention and mapping anticodon statistics to their corresponding codons. 

In practice, tAILoss penalizes probability mass assigned to codons associated with low tRNA availability; increasing tAI is expected to improve translation elongation efficiency. We downloaded the raw tRNA data from GtRNAdb \cite{Chan2015-hs}. Currently, we have incorporated tAILoss parameters for 6 common species, and additional species can be added by providing their tRNA copy-number profiles.

The tAILoss parameters form a vector: $\vec{tAIparam-raw}$. The original tAI computation combines tRNA counts (and their logarithms) with codon counts, which makes the scale dependent on sequence length and complicates normalization. 

\begin{footnotesize}
  \begin{equation}
\begin{blockarray}{cccccccccc}
&UGU&UGC&CAU&CAC&AAA&AAG&UUU&UUC&NNN\\
\begin{block}{c[ccccccccc]}
\vec{tAIparam-raw}= & 2.36 & 4 &  4.72 & 8&  8 & 16.56  & 11. & 6.49 & 0 \\
\end{block}
\end{blockarray}
\end{equation}
\end{footnotesize}

Therefore, analogous to CAILoss, we normalize the parameters within each synonymous codon set by dividing by the maximum value to form a vector $\vec{tAIparam}$:

\begin{footnotesize}
    \begin{equation}
\begin{blockarray}{cccccccccc}
&UGU&UGC&CAU&CAC&AAA&AAG&UUU&UUC&NNN\\
\begin{block}{c[ccccccccc]}
\vec{tAIparam}= & 0.59 & 1 & 0.59 & 1 & 0.48 & 1 & 0.59 & 1 & 0 \\
\end{block}
\end{blockarray}
\end{equation}
\end{footnotesize}

We further set the start codon and stop codon parameters to 1.0 to prevent the model from altering these special tokens. The calculation of tAILoss for input above can be written as:

\begin{footnotesize}
\begin{equation}
  \begin{aligned}
L&=exp(\textbf{O} \cdot log(\vec{tAIparam})^T) \\
&=
\begin{blockarray}{ccccccccc}
  \begin{block}{[ccccccccc]}
   0 & 0.1 & 0.1 & 0.1 & 0.3 & 0.2 & 0 & 0 & 0.2\\
   0 & 0.2 & 0.1 & 0.1 & 0.3 & 0.1 & 0 & 0 & 0.2\\
   0 & 0.1 & 0.2 & 0.1 & 0.3 & 0.2 & 0 & 0 & 0.1\\
   0 & 0.1 & 0.1 & 0.3 & 0.1 & 0.2 & 0 & 0 & 0.2\\
  \end{block}
  \end{blockarray}
\cdot
\begin{blockarray}{c}
  \begin{block}{[c]}
  0.59 \\ 1 \\ 0.59 \\ 1 \\ 0.48 \\ 1 \\ 0.59 \\ 1 \\ 0 \\
  \end{block}
  \end{blockarray}
=
\begin{blockarray}{c}
\begin{block}{[c]}
 0.7610 \\
 0.7610\\
 0.7223\\
 0.8815\\
\end{block}
\end{blockarray}
,\\
&tAILoss=\sum^N_i(1-L_i)
\end{aligned}
\end{equation}
\end{footnotesize}

Eq.~\eqref{eq:tailoss1} summarizes the tAILoss computation, where $\vec{tAIparam}$ denotes the tAI parameter vector. 

\begin{equation}
L=exp(\textbf{O} \cdot log(\vec{tAIparam}^T)) =\begin{blockarray}{c}
\begin{block}{[c]}
  L_{tAI,1}\\
  L_{tAI,2}\\
  ...\\
  L_{tAI,n}\\
\end{block}
\end{blockarray}
,tAILoss=\sum^N_i(1-L_{tAI,i})
\label{eq:tailoss1}
\end{equation}

\subsubsection{MFELoss: Full Sequence Optimization for MFE  }
\label{secmfe}
To improve the minimum free energy of mRNA sequences, we designed the MFELoss to incorporate the knowledge of mRNA stability and half-life into the RNop model. Generally, the lower the MFE of mRNA sequences is, the more stable the mRNA sequences are, and the longer their half-life is, which indicates better expression time and amount in the cell\cite{MATHEWS1999911,david2004}. 

To calculate the MFE of mRNA sequences, there are methods using algorithms which are the $O(N^3)$ time complexity \cite{10.1073/pnas.0401799101, Zhang2023}. These methods can hardly be integrated into neural networks due to time issues--their time consumption is too high for deep learning iterations. Therefore, to make MFE computation fast and utilizable in deep learning, we build a small loss network specialized to predict the MFE of mRNA sequences. 

First, we calculate the MFE value in kJ/mol using RNAStructure\cite{Reuter2010} for all RNA sequences in our collected datasets. 
A very small Multilayer Perceptron (MLP) neural network is applied for rapid estimation of mRNA sequences. We tried different model sizes and found that a small model with 12 layers and 5.4M parameters can achieve high precision, up to $R^2=0.9983$ on the test set. The calculation of MFELoss is shown in the Eq.\ref{eqmfeloss}. The $MFE_{i,gt}$ represents the ground truth MFE of the sequence $i$, and $OP_i$ represents the $i$th output of the optimization model. The $F_{mfe}$ means the $OP_i$ is processed by the MFE model mentioned above. The $\lambda \in [1,+\infty)$ is the MFE ratio to ensure the optimization target is higher than the original MFE. Consequently, the loss operation $Loss$, which can be selected from L1, SmoothL1, MSE, and Charbonnier loss, is applied to these vectors.

\begin{equation}
\label{eqmfeloss}
MFELoss=Loss(\lambda \times MFE_{i,gt},F_{mfe}(Seq_i)) 
\end{equation}

The MFELoss makes the optimization model prone to generating mRNA sequences with better (lower) MFE. Meanwhile, the MFE deviation values of the MFELoss are usually in the $10^2$ order of magnitude, while others are within $(0,1]$. Therefore, the default loss weight of the MFELoss is set to $10^{-2}$.

\subsubsection{Implementation Details}   
During training, the RNop model is applied with PyTorch version 2.1 and trained with CUDA version 12.2. We train each model for 100,000 iterations (equivalently 10 epochs) with a batch size of 64 and a learning rate of $10^{-4}$ with cosine decay. The AdamW~\cite{kingma2014adam,loshchilov2019} optimizer is applied with a 1000 iterations warm-up. All experiments are trained using a single NVIDIA GeForce A800 GPU, running for 18 hours to complete the training process. The models are trained on the training dataset and validated on the validation dataset during the training process, and the test dataset is independent in the subsequent test process. 

\subsubsection{Quantitative Metrics}

We evaluate optimized sequences using three biological metrics (CAI, tAI, MFE) and a set of error-based metrics that quantify fidelity relative to the original sequences.

The CAI is a widely used measure of how well codon usage in an mRNA matches the codon-usage bias of a target species. We compute CAI for both original and optimized sequences to assess their codon optimality in the target species, following \cite{10.1093/nar/15.3.1281,genscript_codon_table}.

\begin{equation}
CAI = f_{CAI}(mRNA Sequence).
\label{eq:cai}
\end{equation}

The tAI evaluates how well an mRNA sequence matches the tRNA pool of a species via tRNA anticodon counts. We compute tAI for the output sequences to quantify their tRNA affinity in the target species, following \cite{tai1,tai2}.

\begin{equation}
tAI = f_{tAI}(mRNA Sequence).
\label{eq:tai}
\end{equation}

The MFE is a thermodynamic descriptor of mRNA structure; lower MFE has been associated with increased stability and longer half-lives \cite{MATHEWS1999911, david2004}. We use RNAStructure \cite{Reuter2010} to compute the MFE of optimized mRNA sequences.

\begin{equation}
MFE = f_{MFE}(mRNA Sequence).
\label{eq:mfe}
\end{equation}

To characterize discrepancies between original and optimized RNA sequences, we further define four error metrics: {length error rate (LER)}, {codon error rate (CER)}, {codon error percentage (CEP)}, and {sequence difference percentage (Sequence Difference)}.

The \textbf{length error rate (LER)} is the fraction of sequences whose optimized length differs from the original (indicating missing or extra codons, which are common in probabilistic models). Let $N$ be the total number of sequences and $N_{LE}$ the number with length errors; then LER is given in Eq.~\ref{eq:LER}.
\begin{small}
\begin{equation}
LER = \frac{N_{LE}}{N}.
\label{eq:LER}
\end{equation}
\end{small}

The \textbf{codon error rate (CER)} is the proportion of sequences that contain at least one incorrect codon, and the \textbf{codon error percentage (CEP)} is the average fraction of incorrect codons among those erroneous sequences. Let $N_{CE}$ (out of $N$ total sequences) denote the number of sequences with codons that do not preserve the original amino-acid sequence, and let $ce_i$ of $n_i$ codons be wrong in the $i$-th such sequence for $i \in [1, N_{CE}]$. CER and CEP are defined in Eqs.~\ref{eq:CER} and \ref{eq:CEP}.
\begin{small}
\begin{equation}
CER = \frac{N_{CE}}{N}.
\label{eq:CER}
\end{equation}
\end{small}

\begin{equation}
CEP = \frac{\sum^{N_{CE}}_{i}{ce_{i}}}{\sum^{N_{CE}}_{i}{n_{i}}}.
\label{eq:CEP}
\end{equation}

The \textbf{sequence difference percentage (Sequence Difference)} measures how strongly error-free sequences are altered at the codon level. Let $N_{NE}$ be the number of sequences without length or codon errors, so that $N = N_{LE} + N_{CE} + N_{NE}$. For each such sequence $j \in [1, N_{NE}]$, let $sd_j$ of $n_j$ codons differ from the original while remaining correct (synonymous). Sequence Difference is defined in Eq.~\ref{eq:SDP}.
\begin{equation}
Sequence Difference = \frac{\sum^{N_{NE}}_{j}{sd_{j}}}{\sum^{N_{NE}}_{j}{n_{j}}}.
\label{eq:SDP}
\end{equation}

When LER is nonzero, CER, CEP, and Sequence Difference are computed only over sequences without length errors, since differing lengths make a codon-wise comparison ill-defined.
\section{Results}

\subsection{RNop Simultaneously Achieves Multi-Objective Optimization and Sequence Fidelity \textit{in Silico}}

In the introduction, the trade-off between \textbf{sequence fidelity}, \textbf{multi-objective optimization}, and \textbf{computational efficiency} was defined as the "impossible triangle". We first sought to validate the ability of RNop to conquer the first two of these challenges through large-scale \textit{in silico} experiments. 

Our first experiment was designed to evaluate RNop's ability to achieve multi-objective optimization in biological metrics while strictly maintaining sequence fidelity. We trained RNop models on the training dataset with over 5,400,000 sequences, targeting different species: "E. coli", "Mouse", "Chinese hamster", and "Human". These models were then benchmarked on a test set of 540,000 sequences. The dataset is preprocessed to avoid data leakage, so all sequences are "unseen" for pretrained models. The calculation process adheres to the methods outlined in the original papers\cite{10.1093/nar/15.3.1281,tai1,tai2}.

\begin{figure}[!htbp]
  \centering
  \includegraphics[width=0.85\linewidth]{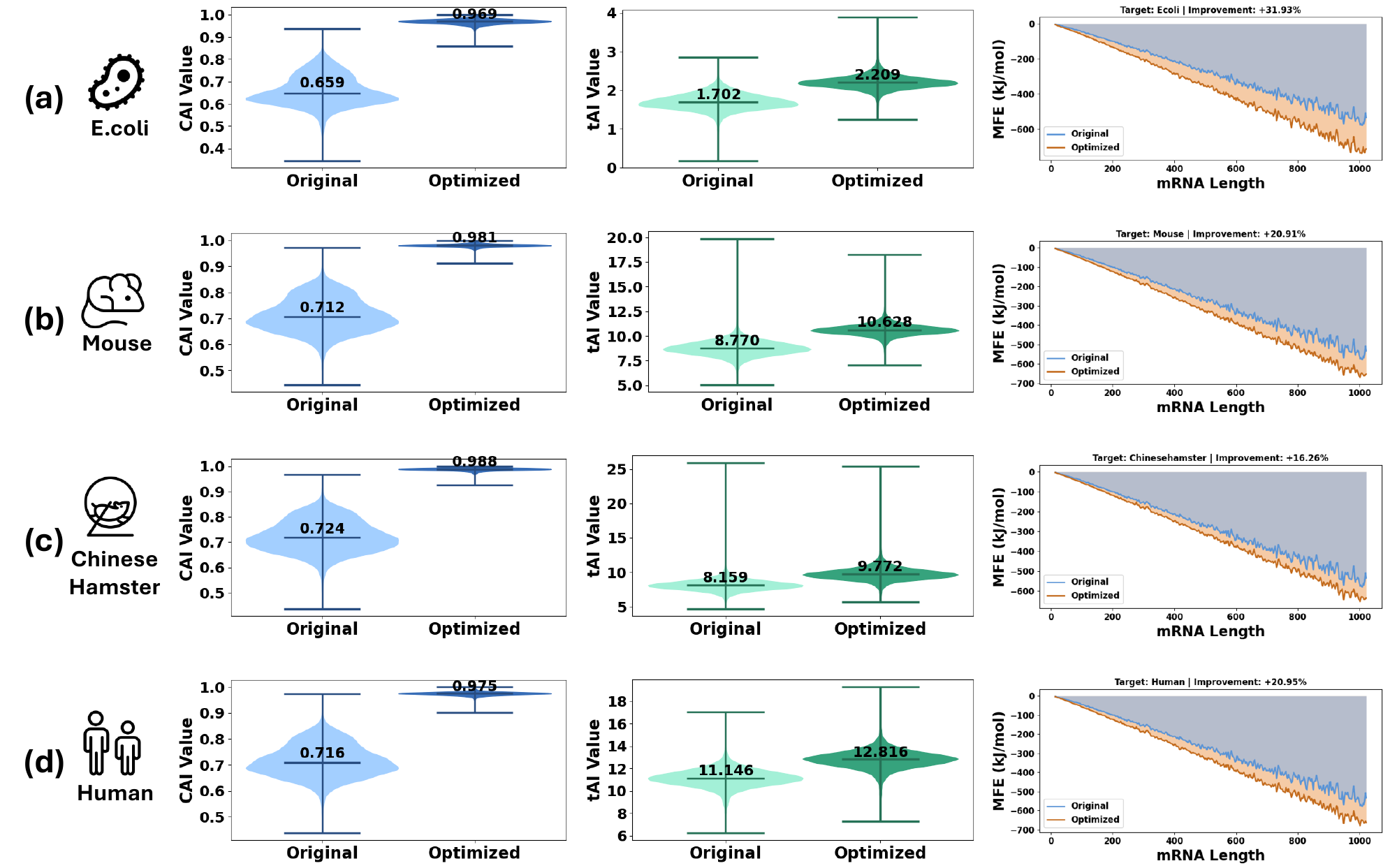}
  \caption{\textbf{The histogram of overall CAI, tAI, and MFE comparison between original and optimized mRNA sequences.} We applied the RNop model to the test dataset with different target species ((a) E. coli, (b) mouse, (c) Chinese hamster, (d) human). No sequence length or codon error was observed. The optimized mRNA sequences have higher overall CAI, tAI, and MFE values, which indicate better codon adaptiveness, tRNA affinity, and chemical stability in the target species.}
  \label{fig:combined_figure_overall}
\end{figure}

The validation results confirm that RNop successfully achieves perfect sequence fidelity, producing \textbf{no undesired amino acid mutations} across the test set and validation set. Meanwhile, the model demonstrated consistent \textbf{multi-objective optimization}. As visualized in the violin plots in Fig.~\ref{fig:combined_figure_overall}, sequences optimized by RNop exhibited significantly higher overall CAI and tAI values and significantly lower MFE values compared to the original across all four parallel experiments. This proved RNop's capacity to simultaneously optimize multiple biological objectives targeting various species.

\begin{figure}[!htbp]
  \centering
  \includegraphics[width=0.85\linewidth]{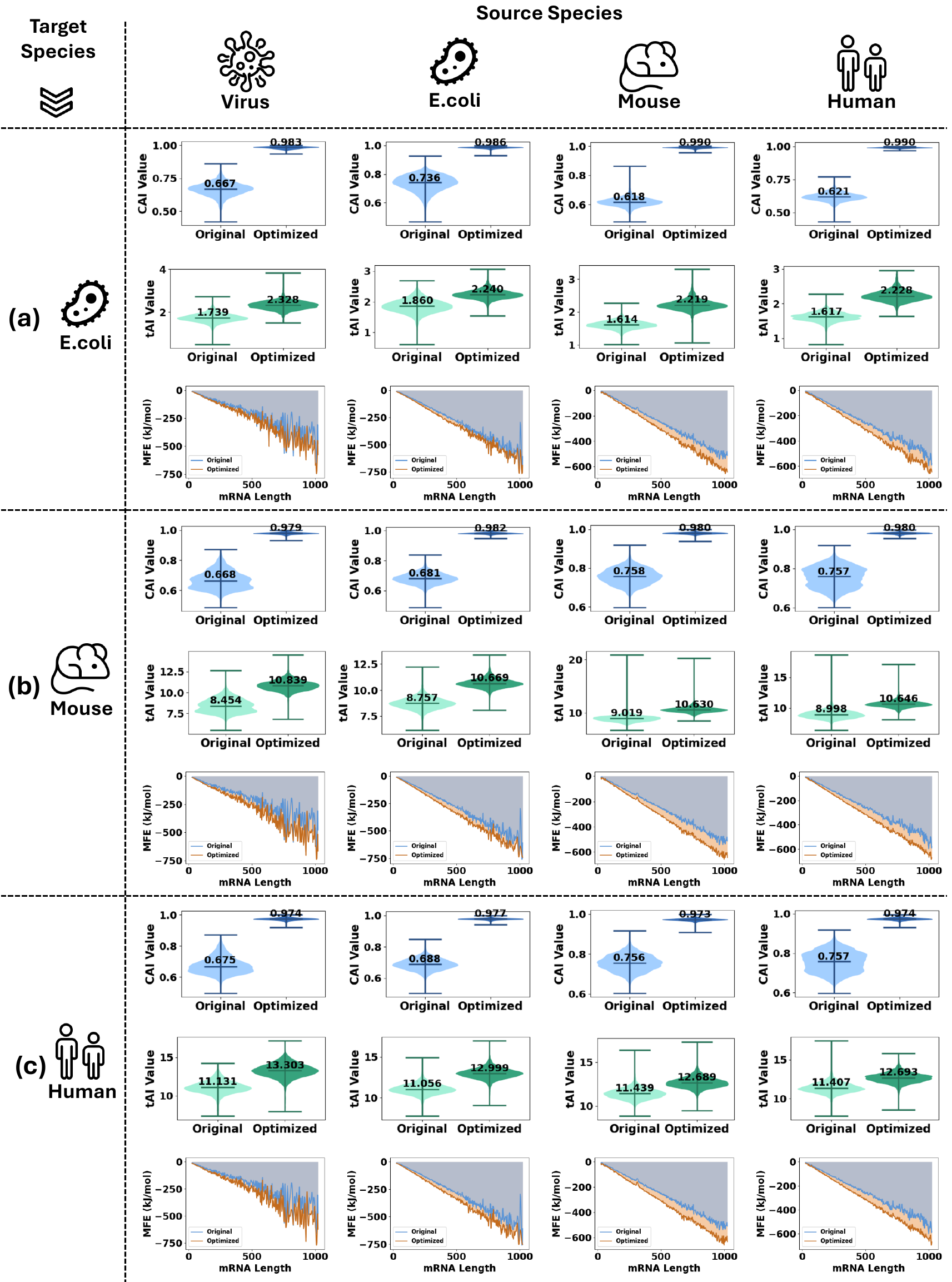}
  \caption{\textbf{The histogram of CAI, tAI, and MFE comparison between mRNA sequences from different species} No sequence length or codon error was observed. (a) The optimized mRNA sequences have better overall CAI, tAI, and MFE values. The data resources are virus, E. coli, mouse, and human, respectively. And the target species is "E. coli". Both the heterologous and homologous sequences are optimized. (b) The target species is "Mouse". (c) The target species is "Human".}
  \label{fig:combined_figure_species}
\end{figure}

Next, we substantiated that our knowledge-infusion framework was genuinely effective in the second \textit{in silico} experiment. A model that truly learned the implicit rules of optimization should not only optimize sequences from heterologous species but also further improve homologous sequences. Therefore, we used the species-specified test set, and the results are shown in Fig.~\ref{fig:combined_figure_species}. When the target species was set to "E. coli", RNop improved metrics for sequences from all species, including virus, mouse, human, and E.coli homologous sequences. This is also consistent for models targeting "Mouse" (Fig.~\ref{fig:combined_figure_species} (b)) and "Human" (Fig.~\ref{fig:combined_figure_species} (c)).

These results substantiate the generalization ability of the RNop model. It shows the advantage of the knowledge-infusion schema over other methods that learn directly from natural sequences. RNop is not memorizing or imitating natural sequences from a specific species; rather, it has learned the explicit biological prior knowledge for that species. Therefore, RNop can accept a sequence from any resource and exhibit robust performance in both heterologous and homologous optimization.

\subsection{RNop provides high-throughput mRNA optimization}

The computation efficiency is another crucial piece of the "impossible triangle" of mRNA optimization. Current optimization methods, such as dynamic programming (DP), have high time complexity, $O(n^3)$ \cite{Terai2016,cohen2003}, leading to burdensome runtimes for sequences longer than 2,000 nucleotides s (nt). Even faster approaches like LinearDesign \cite{Zhang2023} have $O(n^2)$ complexity, and their processing time typically becomes prominent for sequences exceeding 3,000 nt. 

By contrast, RNop own innate advantages for its deep learning architecture in rapid inference and batch processing. The processing time of RNop is tied to the parameter amount and model depth configured at training and does not change afterward. For an input length within the model’s capacity, sequences will be processed in essentially the same time. Meanwhile, the batch processing capability enables RNop to process multiple sequences in parallel. Consequently, RNop can offer a strong advantage in most scenarios due to low per-sequence latency and efficient batch processing.

We evaluate inference time across RNop models with different embedding dimensions, depths, and sequence lengths. The model is running on one Nvidia A800 GPU, with CUDA version 12.2. The results are shown in Fig.\ref{fig:charts} (b). For comparison, CDSFold \cite{Terai2016} requires about 1 hour to design a 2700 nt CDS, and LinearDesign takes minutes to optimize a 2048 nt mRNA, and RNop models achieve throughput up to 44.49 sequences per second for any sequence under 3072 nt. 
Though larger models handling longer sequences need more time, they also achieve better performance than the aforementioned methods.
The higher processing speed enables high-throughput mRNA design and facilitates integration with protein structure and function prediction models. 

\begin{figure}[htbp]
  \centering
  \includegraphics[width=0.85\linewidth]{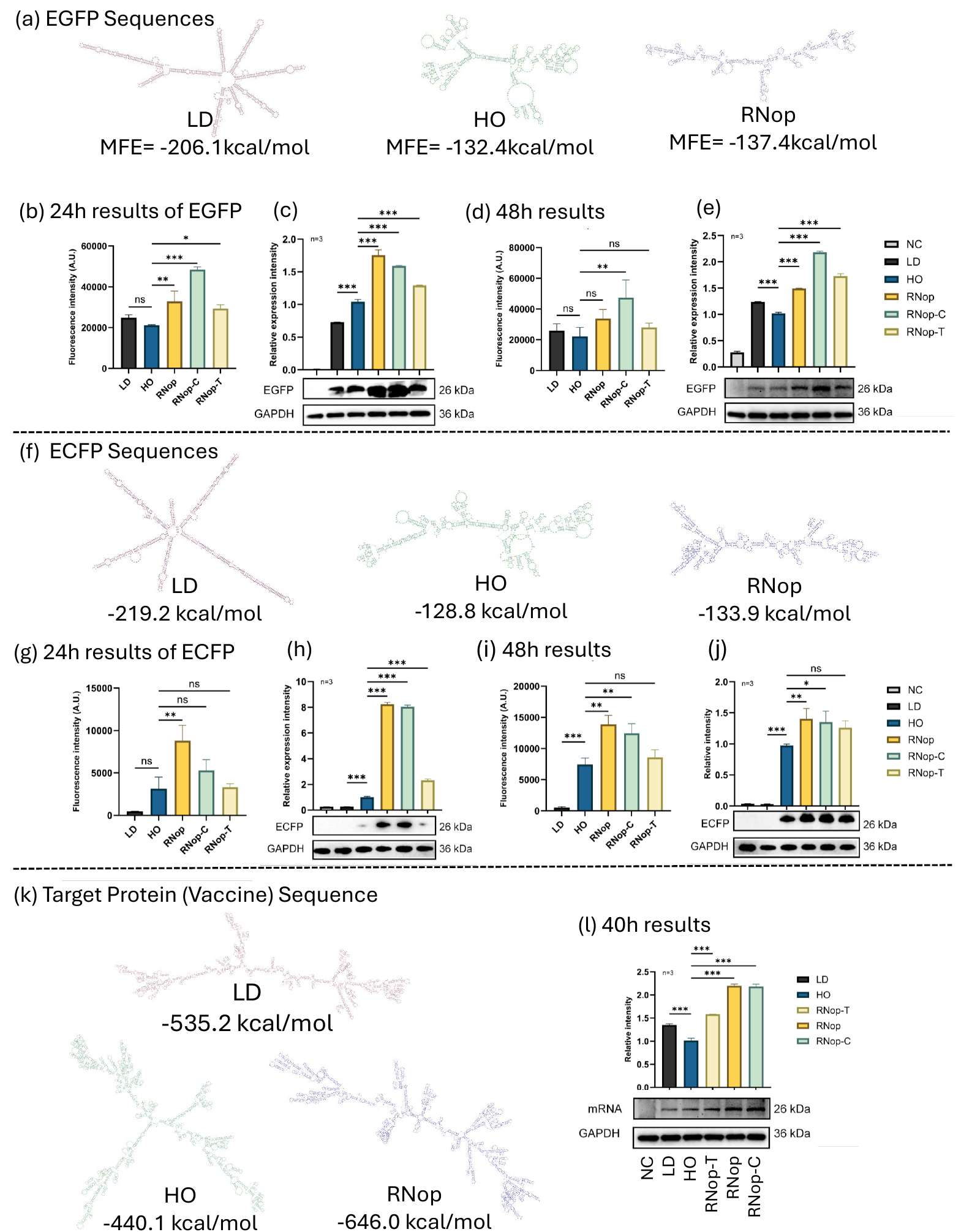}
  \caption{\textbf{\textit{In vitro} analysis of mRNA sequence optimization by the RNop model.} We applied the fluorescent enzyme-labeled instrument and western blot to measure the expression rate of mRNA sequences and summarized these results through densitometric analysis. The Glyceraldehyde-3-phosphate Dehydrogenase (labeled "GAPDH" in the chart) was selected as trimimg standard for cell amounts.
  (a) The visualization of optimized EGFP sequences. Fluorescent and western blot results of EGFP expression at 24 h and 48 h after mRNA transfection are summarized in (b), (d), and (c), (e); (f) The visualization of ECFP. Results at 24 h and 48 h after mRNA transfection are summarized in (g), (i), and (h), (j); The visualization (k) and expression results (l) of the target protein. Replications $n=3, *p<0.05, **p<0.01, ***p<0.001$. }
  \label{figBR3}
\end{figure}

\subsection{Knowledge-Infused Loss Functions Provides Interpretable Control}

Having addressed the "impossible triangle" and validated RNop in biological experiments, we sought deeper insights into the functions of each type of loss function about how they contribute to the optimization process. This exploration is demonstrated by the models with different weights of loss functions, through which we can observe the effect of each loss function on the target sequences.

\textbf{Controllable Sequence Fidelity of GPLoss}

The GPLoss is designed to avoid undesired amino acid (aa) mutation. The mutation here means the protein aa sequences before and after optimization are not synonymous. Without constraint, most deep learning and neural network models may cause mutations, as they are probabilistic models. Some deep learning methods \cite{Fu2020} replace the mutated codons with the original ones to avoid mutation, but this may cause the loss of optimality. Therefore, we designed the GPLoss to avoid mutation generation without post-processing. 

\begin{figure}[htbp]
  \centering
  \includegraphics[width=0.90\linewidth]{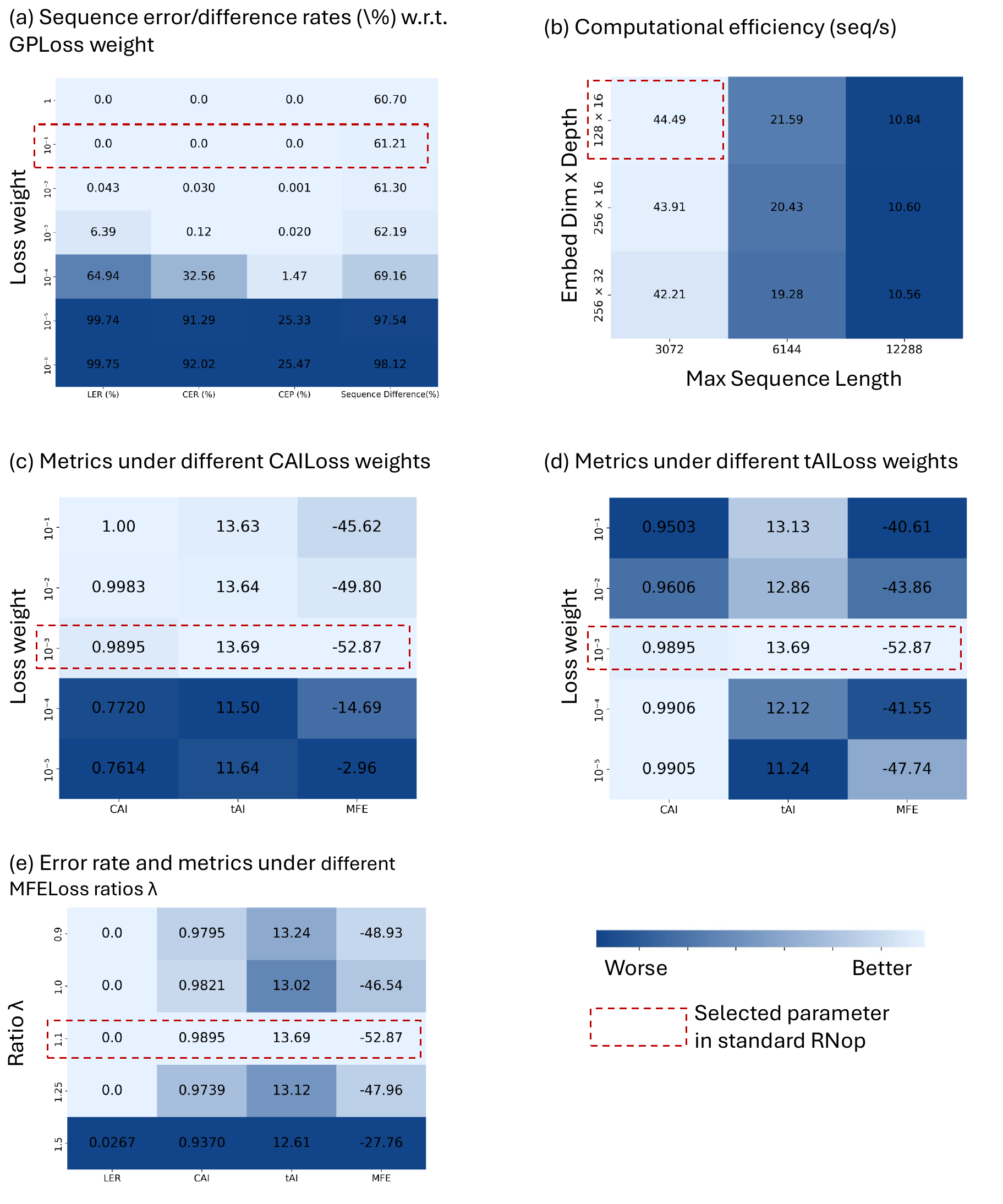}
  \caption{\textbf{The results of throughput and interpretability.}  (a) Results of sequence error/difference rates (\%) w.r.t. GPLoss weight, with setting target species "Human". $10^{-1}$ is enough for synonymous retention; (b) Computational efficiency (sequences per second) under different combinations of embedding dimension, depth of RNop, and max sequence length; (c) CAI, tAI, MFE under different CAILoss weights. Properly setting the loss weight can obtain sequences with best overall metrics; (d)CAI, tAI, MFE under different tAILoss weights; (e)LER, CAI, tAI, MFE under different MFELoss ratios $\lambda$. Setting $\lambda$ too high resulted in unconvergence, while setting it too low resulted in ignoring the MFE optimization. }
  \label{fig:charts}
\end{figure}

We conducted experiments on the RNop model with different GPLoss weights ranging from $1$ to $10^{-6}$ to showcase its controllable flexibility. We can find that the mutation rate is actually adjustable in the RNop model, and it varies with the GPLoss weight. The results, summarized in Fig.\ref{fig:charts} (a), show that high loss weights preserve the synonymous protein sequences after optimization. For weights $\geq 10^{-2}$, the mutation rate is tightly controlled, and neither codon errors nor length errors occur. At the same time, GPLoss offers controllable flexibility: by tuning the loss weight, the mutation rate of optimized sequences can be adjusted from 0\% to nearly 100\%. As the loss weight decreases, the mutation rate increases; at $10^{-6}$, widespread mutations are observed in the output sequences.

These results can explain the role of GPLoss in regulating mutation behavior in the RNop models.

\textbf{Modulation of Codon Usage by CAILoss and tAILoss}

Biologically, the codon usage is not arbitrary in different species. To manifest this preference, the CAI\cite{10.1093/nar/15.3.1281} is proposed to measure the codon usage bias. Meanwhile, the tAILoss specifically focuses on tRNA counts across different species, where the anticodons associated with the highest tRNA counts are considered the most preferred codons. The CAILoss and tAILoss are supposed to regulate the CAI and tAI of output sequences. 

We conducted experiments adjusting the loss weight of CAILoss and tAILoss from $10^{-1}$ to $10^{-5}$. In these experiments, CAILoss and tAILoss were independently evaluated. The results are presented in Fig.\ref{fig:charts} (c) and (d). Mostly, the change in the loss weight of CAILoss leads to a slight corresponding change in the CAI values of output sequences. 
However, though tAI manifests a reduction trend when the loss weight of tAILoss is reduced, when the loss weight is set to $10^{-2}$ or $10^{-1}$, the increase in the loss weight for tAILoss leads to overfitting in some experiment runs.

This trend can explicitly prove that adjusting the loss weight of CAILoss and tAILoss during training can regulate the CAI and tAI of output sequences. Meanwhile, we also note that the loss weights shall balance the contribution of CAILoss and tAILoss. Exaggerating the loss weight of one loss function may cause overfitting or unstable convergence.

\begin{figure}[h!]
  \centering
  \includegraphics[width=0.99\linewidth]{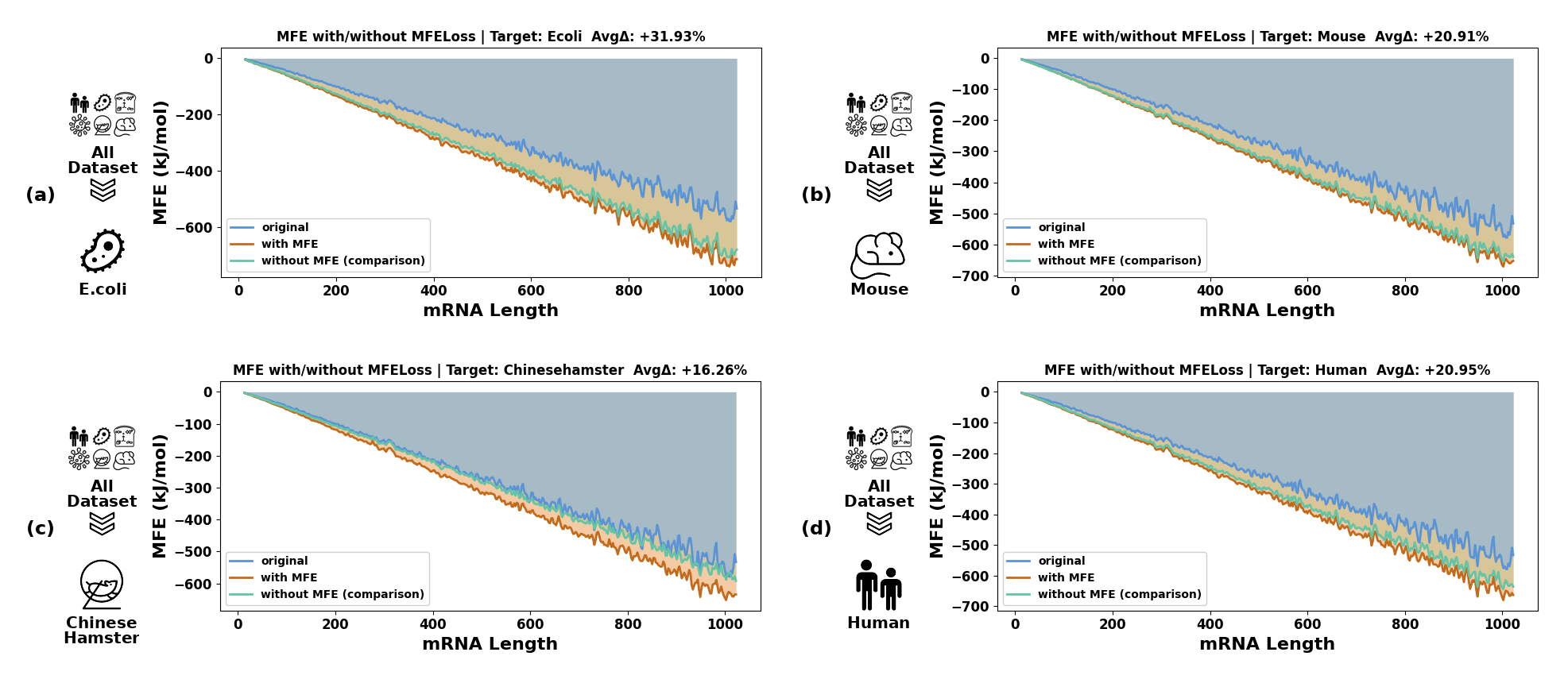}
  \caption{\textbf{The histogram of MFE comparison between sequences original, optimized, and optimized without MFELoss.}  (a)Results of setting target species "E. Coli"; (b)Results of setting target species "Mouse"; (c) Results of setting target species "Chinese hamster"; (d)Results of setting target species "Human". No incorrect sequence length or codon error was observed. The optimized mRNA sequences have lower MFE values than the original sequences. Meanwhile, they also surpass the optimization results from the RNop model with the MFELoss disabled. These experiments demonstrate that the MFELoss works effectively in improving the MFE of mRNA sequences.}
  \label{fig:combined_figure_mfe}
\end{figure}

\textbf{Explicit Optimization of MFE by MFELoss}

MFE profoundly affects the stability of mRNA and the efficiency of translation. Though improvements in CAI and tAI can also reduce MFE \cite{Presnyak2015-xm}, we conducted experiments to isolate the contribution of MFELoss. We compare the MFE of sequences optimized by RNop with and without MFELoss against the original sequences.

As shown in Figure \ref{fig:combined_figure_mfe}, RNop models with MFELoss produce sequences with lower MFE than both the original sequences and those optimized without MFELoss. This pattern holds across target species (“E. coli,” “Mouse,” “Chinese hamster,” and “Human” ). Notably, sequences optimized without MFELoss also exhibit lower MFE than the originals, consistent with prior findings \cite{Presnyak2015-xm}. These results demonstrate that MFELoss effectively improves the MFE of mRNA sequences beyond gains attributable to CAI/tAI alone.

We also introduce a parameter $\lambda$ to ensure the optimization target is higher than the original MFE by setting a target MFE lower than the original. 
The higher ratio $\lambda$ can set the lower target MFE. We conducted experiments on models with $\lambda$ values from 0.9 to 1.5, and the results are summarized in Fig.\ref{fig:charts} (e).

We observe that $\lambda=1.1$ provides the best balance, yielding the largest MFE improvement alongside strong performance on other metrics. When $\lambda$ is too high, low MFE is overemphasized: the target may exceed the maximum attainable improvement for sequences, leading to persistent MFELoss pressure that competes with other objectives and, thus, causes length or codon errors (as seen for $\lambda=1.5$). Conversely, when $\lambda$ is too low, the target is easily satisfied, and the MFE will not be optimized. As a result, sequences exhibit relatively higher MFE. Therefore, we can conclude that the MFELoss ratio $\lambda$ is an important hyperparameter that shall be set within the maximum optimizable capability for sequences to balance the MFE improvement and model convergence. 

\subsection{Sequences optimized by RNop model exhibit higher expression efficiency in \textit{in vitro} experiments}

Our \textit{in silico} experiments demonstrated that RNop successfully addresses the "impossible triangle". However, the golden standard of mRNA optimization is the protein expression rate within a complex biological environment. We conducted biological experiments to demonstrate the \textit{in vitro} effectiveness of RNop and confirm its ability to preserve protein function.

We selected two of the most widely used and highly optimized reporter proteins: the enhanced green fluorescent protein (EGFP) and the enhanced cyan fluorescent protein (ECFP). As these sequences are already well-studied and optimized, any significant improvement is exceptionally difficult to achieve. Meanwhile, a sequence of a key antigen protein, representing a potential component for mRNA vaccine production, was selected as the target protein to bridge the gap between model proteins and real-world applications in the final experiments.

The optimized sequences were compared against control groups: The tested mRNA sequences were labeled as: NC (negative control), HO (original Human-Optimized), LD (LinearDesign) \cite{Zhang2023}, and three variants from our RNop model grid with different loss weights: RNop (Standard RNop model); RNop-C (RNop model optimized with a 10-fold increase in CAILoss weight); and RNop-T (RNop model optimized with a 10-fold increase in tAILoss weight). These sequences are transfected and then expressed \textit{in vitro} with the methods explained in the Section.Method. The expression rate was measured through the fluorescent enzyme-labeled instrument and western blot at 24 hours and 48 hours. The number of total replications is 3, and the significance levels are $*p<0.05, **p<0.01, ***p<0.001$.

\textbf{Results of EGFP optimization demonstrated the effectiveness of RNop:}

In the first experiment, we use the most widely used fluorescence protein, the EGFP. The results are presented in Fig.\ref{figBR3} (a-e).

The fluorescence measurement results (Fig.\ref{figBR3} (b,d)) show a clear advantage of RNop over other groups. At both 24 and 48 hours post-transfection, variants of RNop consistently achieved the highest fluorescence intensity, indicating high protein expression levels. Its performance significantly surpassed the compared groups. The RNop-C variant is the best optimization model for EGFP; its optimized sequence yielded up to 2.28-fold higher expression than the HO group at 48 hours. Through the western blot results (Fig.\ref{figBR3} (c,e)), we can draw a similar conclusion, with RNop variants yielding higher expression than LD and HO. These results demonstrate RNop's ability to obtain substantial performance gains even from already well-optimized sequences.

\textbf{Results of ECFP optimization showed the robustness:}

We extended the validation to the ECFP, and the results are shown in Fig.\ref{figBR3} (f-j).

RNop variants again outperformed other groups significantly, exhibiting substantial increases relative to the original HO sequence. Meanwhile, the LinearDesign optimized sequence suffered an expression failure: protein expression and fluorescence levels were near 0 or near the NC group at both 24 and 48 hours. 

The failure in the LinearDesign group can showcase the concept of the "impossible triangle." When optimizing CAI and MFE, LinearDesign may have introduced excessive secondary structure modifications. Visualization analysis suggests the ECFP sequence was modified into a structure with nearly all nucleotides paired. This structure may hinder translation initiation or elongation and ultimately cause expression failure. Conversely, the multiple optimization objectives of the RNop framework prevent the domination and over-emphasis of any single factor, and thus avoid the expression failure. This highlighted the critical role of sequence fidelity and balanced multi-objective optimization, which are the key strengths of the RNop framework.

\textbf{Optimization of target protein demonstrated the versatility :}

Then, we applied RNop to optimize the sequence of a key antigen protein, a component of an mRNA vaccine. This experiment can demonstrate the effectiveness of RNop on a sequence directly relevant to therapeutic and industrial goals.

The results of the target protein are shown in Figure \ref{figBR3} (k,l). In these experiments, all optimization methods yielded significant expression increases compared to the original. The standard RNop once again achieved the highest protein expression levels at 40 hours post-transfection, outperforming LD. Computational analysis indicated that the optimized sequence of RNop and LD has moderate structural changes. The similar structure with HO permitted the protein expression, yet the predicted MFE of the LD sequence suggested insufficient optimization in MFE compared to the sequences optimized by our method. 

In conclusion, these \textit{in vitro} experiments, including highly optimized reporters and a therapeutically relevant antigen, validate the effectiveness of RNop. RNop models consistently delivered optimized sequences yielding superior protein expression compared to both original sequences and the state-of-the-art LinearDesign algorithm, while maintaining crucial sequence fidelity and avoiding expression failures caused by unbalanced optimization.

\subsection{Summarization: The Controllable Solution of the "Impossible Triangle"}

In summary, the \textit{in silico}, \textit{in vitro}, and loss function analyses demonstrate that RNop successfully resolves the "impossible triangle" situation of current mRNA optimization methods.

The \textit{in silico} experiments confirmed RNop's ability to achieve robust multi-objective optimization while strictly maintaining perfect sequence fidelity. Meanwhile, \textit{in vitro} validation confirmed these computational improvements can translate to real-world protein expression (up to 2.28-fold). Our RNop models also avoided the expression failures associated with unbalanced optimization and outperformed SOTA methods. Furthermore, RNop accomplishes all of these while achieving high computational efficiency, enabling high-throughput applications.

Furthermore, the loss function analyses provide the explicit explanation for the controllability of RNop: the K-iL framework is not a "black box". Each loss function provides interpretable and tunable control over its targeted biological objective. After having validated RNop's performance and mechanism, we will now discuss the broader implications of these findings.
\section{Discussion}

In the \textit{in vitro} experiments, the expression failure of LinearDesign optimized ECFP is an example explaining why multi-factor optimization is important. The overemphasized optimization factor would lead to fatal failure. However, the RNop framework avoided this problem with its balanced and biological knowledge-infused approach. Meanwhile, the sequence fidelity has been a problem for probabilistic models, while the RNop framework achieves zero mutation rate and codon error, reaching the same level as algorithm methods. These advantages, working with the innate advantages of deep learning: batch processing and rapid inference, make RNop outstanding among current deep-learning-based methods.

\textbf{Controllable synonymous retention by K-iL framework.}
Many deep-learning-based codon optimization pipelines handle protein mutations via post-processing: after decoding, non-synonymous codons are replaced by the original codons to restore the amino-acid sequence \cite{Fu2020}. While this can remove mutations, it breaks the optimized distribution and may undo improvements in other optimization objectives. In contrast, RNop enforces synonymy during optimization: GPLoss imposes a differentiable, matrix-based constraint directly on the codon probability space. As a result, strict synonymous outputs can be achieved without post-hoc repair, and the strength of this constraint can be continuously tuned via the GPLoss weight---yielding either strict synonymy with high loss weight or controlled flexibility with low loss weight when desired.

\textbf{Knowledge-infused learning: from physics to biology.}
Physics-informed neural networks (PINNs) \cite{raissi2019physics} have demonstrated how encoding physical laws into loss terms can guide neural networks to solve partial differential equations and related inverse problems. RNop extends this knowledge-infused paradigm to biological sequence design: instead of PDE residuals or conservation laws, we encode biological priors such as codon–amino-acid mappings, codon-usage bias, tRNA adaptation, and structural stability as mechanism-aligned loss functions for mRNA optimization: CAILoss, tAILoss, and MFELoss. This cross-domain perspective suggests that K-iL is a general strategy for structuring learning problems in complex scientific domains, not limited to physics but also effective for high-dimensional biological sequence design.

\textbf{Explainability and controllable optimization.}
The explainability is another major problem for mRNA optimization. The former attempts in deep learning based models, such as works of Gong et al.\cite{deepmRNAfullopt2023, doi:10.1126/science.adr8470}, Goulet et al.\cite{Goulet2023}, and Jain et al.\cite{Jain2023}, followed the strategy of learning from sequences of specific species and assumed the natural mRNA sequences contain the best pattern of possible sequences. 
However, they hardly discussed the rational reasons for this approach. Gong et al.\cite{doi:10.1126/science.adr8470} also claimed that the CDS optimization method in the GEMORNA approach that they had proposed is likely to be a black box, and the explainability of this model is limited for the implicitly learned features. 

On the contrary, the knowledge-infused RNop framework provides a more interpretable explanation for the optimization process: all learned features are explicit and can be explained through biological findings. \cite{10.1093/nar/15.3.1281,tai1,tai2,Mathews1997AnUR, 10.1073/pnas.0401799101} The K-iL framework directly addresses the "black box" problem. As our loss function studies demonstrated, the ability to modulate the contribution of each loss function offers a clear view into the optimization process and enables controllable mRNA design. Thus, the RNop framework provides a powerful approach to make deep learning-based mRNA optimization get rid of the limitations of implicit and "black-box" learning from natural sequences.

\textbf{Effects of different backbone architectures on RNop.}
We chose ViT as our cornerstone model, which seems unusual. However, the choice of the ViT model is not an arbitrary decision but a critical, rationally-driven decision to enable our K-iL framework according to experiments. In our exploration of cornerstone architectures for mRNA sequence optimization, we systematically evaluated linear networks\cite{KOTHARI1993119, NEURIPS2021_cba0a4ee}, sequence-to-sequence models\cite{Fu2020, Goulet2023}, Transformer-based language models\cite{devlin2019bert,li2023codonbert,2020t5,dai2019transformerxl}, and Vision Transformers (ViT)\cite{dosovitskiy2021,liu2021swin}. As detailed in Supplementary information, alternative architectures all failed: linear models and Transformers (BERT, T5) consistently suffered from overfitting, while LSTMs were structurally incompatible with full-sequence losses like MFELoss. The ViT architecture uniquely provided the necessary parameter-efficiency and global dependency modeling required to balance all four knowledge-infused losses simultaneously. The ViT functions as an extraordinary foundation for RNop, inspiring us that the success of the K-iL framework depends not just on the prior knowledge, but on a corresponding adequate architecture to support it.

\textbf{Modular framework of RNop enables future incorporation of UTR generation.}
The UTR design is deliberately decoupled from our study on CDS optimization, which is a methodological choice. Unlike the biological priors we have mentioned governing the CDS, UTR efficiency is governed by fundamentally distinct factors like RBP binding sites, IRES elements, and miRNA avoidance. \cite{deepmRNAfullopt2023,liu2024jointdesign5untranslated,LIU2024168804,Chu2024} Logically speaking, we can regard UTR design as a totally different task from CDS optimization. Forcing these two disparate tasks of different biological logic into a single end-to-end model will risk creating a new, more complex "black box." It may also make it impossible to disentangle which factors contribute to the final expression. 

Meanwhile, UTR generation operates at the nucleotide level, using a different vocabulary from the codon-based RNop model. Our coding rules for CDS optimization are designed for codon-level tokenization. Incorporating UTR generation would require expanding the vocabulary to single nucleotides, potentially introducing instability in the current framework. Also, to preserve simplicity and efficiency, we prefer to separate UTR generation from the RNop implementation.

Our K-iL approach, by its very nature, demands modularity and interpretability. By separating these components, we ensure that the optimization of the CDS is mechanistically transparent. This modularity is an advantage for integrating future works. It creates a potential pathway to develop a separate compatible K-iL module for UTRs. The methodology applied in RNop allows us to rationally combine an optimized CDS with optimized UTR components, from 5' UTR to poly(A) tail, rather than relying on a whole but "black box" optimization.

\textbf{RNop is ready for real world protein production.}
The experiments on the target antigen protein for mRNA vaccine also provide a meaningful insight into the potential of RNop models in protein production. The mRNA vaccine component can be a perfect representation of all other proteins, not only in mRNA vaccines, but also the proteins in industry, medicine, and laboratory that require high expression in target species. RNop enables the optimization across a large scope of species, thus, the protein expression system is not limited to human cells. It can boost the production through yeast and E. coli, the expression in Chinese Hamster Ovary (CHO) cells, and other systems that can be utilized in protein production. The RNop can work as a pivot for the mRNA optimization that empowers the heterologous protein to be easily expressed from one species to another. Especially, due to the rapid response speed, the RNop can provide optimized sequences with unparalleled low-cost and high-throughput.

\section{conclusion}
In conclusion, we demonstrate that the "impossible triangle" of mRNA optimization—the long-held trade-off between sequence fidelity, multi-objective capability, and computational efficiency—is not a fundamental limit but a solvable engineering challenge. The RNop framework successfully solves this challenge by integrating biological prior knowledge into a Transformer architecture in an extensible and modular design. More importantly, this work represents an impactful shift in mRNA optimization methodology. By making knowledge infusion explicit, controllable, and interpretable, RNop begins to transform mRNA design from a "black-box" operation to a predictable and explainable engineering discipline.

The true potential of RNop lies in the modularity we developed in its conceptualization. The RNop framework is not a static solution but a dynamic platform. New biological insights—such as the Mean Typical Decoding Rate (MTDR) \cite{Dana2015MTDR} or tissue-specific tRNA profiles—can be added as plug-in, mechanism-aligned loss functions, enabling targeted optimization objectives without redesigning the backbone model. In parallel, expanding to extra species is straightforward in this platform setting: species-specific priors like CAI and tAI can be incorporated by updating the corresponding loss-parameter tables. This modularity also supports future integration of a compatible UTR module, allowing more rational end-to-end engineering of complete mRNAs, from 5' UTR to poly(A) tail.

The impact of RNop can extend far beyond mRNA. The core principle of RNop— an approach infusing biological knowledge into deep learning models —provides a powerful template for other complex challenges in computational biology. This strategy can be applied to synthetic genome generation, protein design, and biological dark matter exploration. By demonstrating how to guide deep learning with the clarity of biological knowledge, RNop provides a scalable path to engineering biology rather than exploring in the dark. 

\section*{Acknowledgements}

The authors would like to thank Prof. Sanyang Han and Dr. Hongtian Yang from Tsinghua University, Tsinghua Shenzhen International Graduate School, and Prof. Yan Zhao from Sun Yat-sen University, School of Chemical Engineering and Technology for sharing their valuable experience. 

\paragraph*{Competing interests:} The authors declare no competing interests.
\paragraph*{Author contributions:}
\textbf{Zheng Gong:} Conceptualization, Methodology, Validation, Data Curation, Writing - Original Draft, Visualization. \textbf{Ziyi Jiang:} Conceptualization, Validation, Writing - Review and Editing. \textbf{Weihao Gao:} Conceptualization, Writing - Review and Editing. \textbf{Yuanyuan Wang:} Validation. \textbf{Zhining Cai:} Validation. 
\textbf{Zhuo Deng:} Writing - Review and Editing. \textbf{Lan Ma:} Funding acquisition, Supervision.

\paragraph*{Data statement:}
The data and code are available at \url{https://github.com/HudenJear/RPLoss} for justified usage and research upon request. Please contact the corresponding author if you have any questions about the data and code.

\appendix
\section{My Appendix}
Appendix sections are coded under \verb+\appendix+.

\verb+\printcredits+ command is used after appendix sections to list 
author credit taxonomy contribution roles tagged using \verb+\credit+ 
in frontmatter.

\printcredits

\bibliographystyle{cas-model2-names}

\bibliography{cas-refs}

@article{10.1093/nar/15.3.1281,
    author = {Sharp, Paul M. and Li, Wen-Hsiung},
    title = "{The codon adaptation index-a measure of directional synonymous codon usage bias, and its potential applications}",
    journal = {Nucleic Acids Research},
    volume = {15},
    number = {3},
    pages = {1281-1295},
    year = {1987},
    month = {02},
    issn = {0305-1048},
    doi = {10.1093/nar/15.3.1281},
    url = {https://doi.org/10.1093/nar/15.3.1281},
    eprint = {https://academic.oup.com/nar/article-pdf/15/3/1281/3988268/15-3-1281.pdf},
}

@misc{liu2024jointdesign5untranslated,
      title={Joint Design of 5' Untranslated Region and Coding Sequence of mRNA}, 
      author={Yang Liu and Jie Gao and Xiaonan Zhang and Xiaomin Fang},
      year={2024},
      eprint={2410.20781},
      archivePrefix={arXiv},
      primaryClass={q-bio.BM},
      url={https://arxiv.org/abs/2410.20781}, 
}

@article{LIU2024168804,
title = {Rational Design of Untranslated Regions to Enhance Gene Expression},
journal = {Journal of Molecular Biology},
volume = {436},
number = {22},
pages = {168804},
year = {2024},
issn = {0022-2836},
doi = {https://doi.org/10.1016/j.jmb.2024.168804},
url = {https://www.sciencedirect.com/science/article/pii/S0022283624004261},
author = {Mingchun Liu and Zhuoer Jin and Qing Xiang and Huawei He and Yuhan Huang and Mengfei Long and Jicheng Wu and Cheng {Zhi Huang} and Chengde Mao and Hua Zuo}
}

@article{Chu2024,
author={Chu, Yanyi
and Yu, Dan
and Li, Yupeng
and Huang, Kaixuan
and Shen, Yue
and Cong, Le
and Zhang, Jason
and Wang, Mengdi},
title={A 5' UTR language model for decoding untranslated regions of mRNA and function predictions},
journal={Nature Machine Intelligence},
year={2024},
month={Apr},
day={01},
volume={6},
number={4},
pages={449-460},
issn={2522-5839},
doi={10.1038/s42256-024-00823-9},
url={https://doi.org/10.1038/s42256-024-00823-9}
}

@article{
doi:10.1126/science.adr8470,
author = {He Zhang  and Hailong Liu  and Yushan Xu  and Haoran Huang  and Yiming Liu  and Jia Wang  and Yan Qin  and Haiyan Wang  and Lili Ma  and Zhiyuan Xun  and Xuzhuang Hou  and Timothy K. Lu  and Jicong Cao },
title = {Deep generative models design mRNA sequences with enhanced translational capacity and stability},
journal = {Science},
volume = {0},
number = {0},
pages = {eadr8470},
year = {},
doi = {10.1126/science.adr8470},
URL = {https://www.science.org/doi/abs/10.1126/science.adr8470},
eprint = {https://www.science.org/doi/pdf/10.1126/science.adr8470}}

@article {Brixi2025.02.18.638918,
	author = {Brixi, Garyk and Durrant, Matthew G. and Ku, Jerome and Poli, Michael and Brockman, Greg and Chang, Daniel and Gonzalez, Gabriel A. and King, Samuel H. and Li, David B. and Merchant, Aditi T. and Naghipourfar, Mohsen and Nguyen, Eric and Ricci-Tam, Chiara and Romero, David W. and Sun, Gwanggyu and Taghibakshi, Ali and Vorontsov, Anton and Yang, Brandon and Deng, Myra and Gorton, Liv and Nguyen, Nam and Wang, Nicholas K. and Adams, Etowah and Baccus, Stephen A. and Dillmann, Steven and Ermon, Stefano and Guo, Daniel and Ilango, Rajesh and Janik, Ken and Lu, Amy X. and Mehta, Reshma and Mofrad, Mohammad R.K. and Ng, Madelena Y. and Pannu, Jaspreet and R{\'e}, Christopher and Schmok, Jonathan C. and John, John St. and Sullivan, Jeremy and Zhu, Kevin and Zynda, Greg and Balsam, Daniel and Collison, Patrick and Costa, Anthony B. and Hernandez-Boussard, Tina and Ho, Eric and Liu, Ming-Yu and McGrath, Thomas and Powell, Kimberly and Burke, Dave P. and Goodarzi, Hani and Hsu, Patrick D. and Hie, Brian L.},
	title = {Genome modeling and design across all domains of life with Evo 2},
	elocation-id = {2025.02.18.638918},
	year = {2025},
	doi = {10.1101/2025.02.18.638918},
	publisher = {Cold Spring Harbor Laboratory},
	URL = {https://www.biorxiv.org/content/early/2025/02/21/2025.02.18.638918},
	eprint = {https://www.biorxiv.org/content/early/2025/02/21/2025.02.18.638918.full.pdf},
	journal = {bioRxiv}
}

@ARTICLE{Paremskaia2024bn,
  title    = "Codon-optimization in gene therapy: promises, prospects and
              challenges",
  author   = "Paremskaia, Anastasiia Iu and Kogan, Anna A and Murashkina,
              Anastasiia and Naumova, Daria A and Satish, Anakha and Abramov,
              Ivan S and Feoktistova, Sofya G and Mityaeva, Olga N and
              Deviatkin, Andrei A and Volchkov, Pavel Yu",
  journal  = "Front Bioeng Biotechnol",
  volume   =  12,
  pages    = "1371596",
  month    =  mar,
  year     =  2024,
  address  = "Switzerland",
  language = "en"
}

@INPROCEEDINGS{Muhammad2020,
  author={Nabeel Asim, Muhammad and Imran Malik, Muhammad and Dengel, Andreas and Ahmed, Sheraz},
  booktitle={2020 International Joint Conference on Neural Networks (IJCNN)}, 
  title={K-mer Neural Embedding Performance Analysis Using Amino Acid Codons}, 
  year={2020},
  volume={},
  number={},
  pages={1-8},
  doi={10.1109/IJCNN48605.2020.9206892}}

@inproceedings{Nicole2023PredictionCodonBERT,
author = {Babjac, Ashley Nicole and Lu, Zhixiu and Emrich, Scott J},
title = {CodonBERT: Using BERT for Sentiment Analysis to Better Predict Genes with Low Expression},
year = {2023},
isbn = {9798400701269},
publisher = {Association for Computing Machinery},
address = {New York, NY, USA},
url = {https://doi.org/10.1145/3584371.3613013},
doi = {10.1145/3584371.3613013},
booktitle = {Proceedings of the 14th ACM International Conference on Bioinformatics, Computational Biology, and Health Informatics},
articleno = {29},
numpages = {6},
location = {Houston, TX, USA},
series = {BCB '23}
}

@Article{Fu2020,
author={Fu, Hongguang
and Liang, Yanbing
and Zhong, Xiuqin
and Pan, ZhiLing
and Huang, Lei
and Zhang, HaiLin
and Xu, Yang
and Zhou, Wei
and Liu, Zhong},
title={Codon optimization with deep learning to enhance protein expression},
journal={Scientific Reports},
year={2020},
month={Oct},
day={19},
volume={10},
number={1},
pages={17617},
issn={2045-2322},
doi={10.1038/s41598-020-74091-z},
url={https://doi.org/10.1038/s41598-020-74091-z}
}

@article{Dana2015MTDR,
    author = {Dana, Alexandra and Tuller, Tamir},
    title = {Mean of the Typical Decoding Rates: A New Translation Efficiency Index Based on the Analysis of Ribosome Profiling Data},
    journal = {G3 Genes|Genomes|Genetics},
    volume = {5},
    number = {1},
    pages = {73-80},
    year = {2015},
    month = {01},
    issn = {2160-1836},
    doi = {10.1534/g3.114.015099},
    url = {https://doi.org/10.1534/g3.114.015099},
    eprint = {https://academic.oup.com/g3journal/article-pdf/5/1/73/37191150/g3journal0073.pdf},
}

@Article{Jain2023,
author={Jain, Rishab
and Jain, Aditya
and Mauro, Elizabeth
and LeShane, Kevin
and Densmore, Douglas},
title={ICOR: improving codon optimization with recurrent neural networks},
journal={BMC Bioinformatics},
year={2023},
month={Apr},
day={04},
volume={24},
number={1},
pages={132},
issn={1471-2105},
doi={10.1186/s12859-023-05246-8},
url={https://doi.org/10.1186/s12859-023-05246-8}
}

@article{BUHR2016341,
title = {Synonymous Codons Direct Cotranslational Folding toward Different Protein Conformations},
journal = {Molecular Cell},
volume = {61},
number = {3},
pages = {341-351},
year = {2016},
issn = {1097-2765},
doi = {https://doi.org/10.1016/j.molcel.2016.01.008},
url = {https://www.sciencedirect.com/science/article/pii/S1097276516000095},
author = {Florian Buhr and Sujata Jha and Michael Thommen and Joerg Mittelstaet and Felicitas Kutz and Harald Schwalbe and Marina V. Rodnina and Anton A. Komar}
}

@article{Nikita2024mRNAid,
    author = {Vostrosablin, Nikita and Lim, Shuhui and Gopal, Pooja and Brazdilova, Kveta and Parajuli, Sushmita and Wei, Xiaona and Gromek, Anna and Prihoda, David and Spale, Martin and Muzdalo, Anja and Greig, Jamie and Yeo, Constance and Wardyn, Joanna and Mejzlik, Petr and Henry, Brian and Partridge, Anthony W and Bitton, Danny A},
    title = {mRNAid, an open-source platform for therapeutic mRNA design and optimization strategies},
    journal = {NAR Genomics and Bioinformatics},
    volume = {6},
    number = {1},
    pages = {lqae028},
    year = {2024},
    month = {03},
    issn = {2631-9268},
    doi = {10.1093/nargab/lqae028},
    url = {https://doi.org/10.1093/nargab/lqae028},
    eprint = {https://academic.oup.com/nargab/article-pdf/6/1/lqae028/56958324/lqae028.pdf},
}

@article{TANEDA20201811,
title = {COSMO: A dynamic programming algorithm for multicriteria codon optimization},
journal = {Computational and Structural Biotechnology Journal},
volume = {18},
pages = {1811-1818},
year = {2020},
issn = {2001-0370},
doi = {https://doi.org/10.1016/j.csbj.2020.06.035},
url = {https://www.sciencedirect.com/science/article/pii/S200103702030324X},
author = {Akito Taneda and Kiyoshi Asai}
}

@article{Goulet2023,
author = {Goulet, Dennis R. and Yan, Yongqi and Agrawal, Palak and Waight, Andrew B. and Mak, Amanda Nga-sze and Zhu, Yi},
title = {Codon Optimization Using a Recurrent Neural Network},
journal = {Journal of Computational Biology},
volume = {30},
number = {1},
pages = {70-81},
year = {2023},
doi = {10.1089/cmb.2021.0458},
    note ={PMID: 35727687},

URL = { 
    
        https://doi.org/10.1089/cmb.2021.0458
    
    

},
eprint = { 
    
        https://doi.org/10.1089/cmb.2021.0458
    
    

}
}

@article{kursuncu2019knowledge,
  title={Knowledge infused learning (k-il): Towards deep incorporation of knowledge in deep learning},
  author={Kursuncu, Ugur and Gaur, Manas and Sheth, Amit},
  journal={arXiv preprint arXiv:1912.00512},
  year={2019}
}

@article{pandya2022infusedheart,
  title={InfusedHeart: A novel knowledge-infused learning framework for diagnosis of cardiovascular events},
  author={Pandya, Sharnil and Gadekallu, Thippa Reddy and Reddy, Praveen Kumar and Wang, Weizheng and Alazab, Mamoun},
  journal={IEEE Transactions on Computational Social Systems},
  volume={11},
  number={3},
  pages={3060--3069},
  year={2022},
  publisher={IEEE}
}

@article{raissi2019physics,
  title={Physics-informed neural networks: A deep learning framework for solving forward and inverse problems involving nonlinear partial differential equations},
  author={Raissi, Maziar and Perdikaris, Paris and Karniadakis, George E},
  journal={Journal of Computational physics},
  volume={378},
  pages={686--707},
  year={2019},
  publisher={Elsevier}
}

@article{deepmRNAfullopt2023,
    author = {Gong, Haoran and Wen, Jianguo and Luo, Ruihan and Feng, Yuzhou and Guo, JingJing and Fu, Hongguang and Zhou, Xiaobo},
    title = {Integrated mRNA sequence optimization using deep learning},
    journal = {Briefings in Bioinformatics},
    volume = {24},
    number = {1},
    pages = {bbad001},
    year = {2023},
    month = {01},
    issn = {1477-4054},
    doi = {10.1093/bib/bbad001},
    url = {https://doi.org/10.1093/bib/bbad001},
    eprint = {https://academic.oup.com/bib/article-pdf/24/1/bbad001/48782582/bbad001.pdf},
}

@misc{genscript_codon_table,
title = {GenScript Codon Frequency Table},
author   = "GenScript",
howpublished = {\url{https://www.genscript.com/tools/codon-frequency-table}},
note = {Accessed: 2024/07/03},
}

@ARTICLE{tai1,
  title    = "Unexpected correlations between gene expression and codon usage
              bias from microarray data for the whole Escherichia coli {K-12}
              genome",
  author   = "dos Reis, Mario and Wernisch, Lorenz and Savva, Renos",
  journal  = "Nucleic Acids Res",
  volume   =  31,
  number   =  23,
  pages    = "6976--6985",
  month    =  dec,
  year     =  2003,
  address  = "England",
  language = "en"
}

@article{tai2,
    author = {Reis, Mario dos and Savva, Renos and Wernisch, Lorenz},
    title = "{Solving the riddle of codon usage preferences: a test for translational selection}",
    journal = {Nucleic Acids Research},
    volume = {32},
    number = {17},
    pages = {5036-5044},
    year = {2004},
    month = {01},
    issn = {0305-1048},
    doi = {10.1093/nar/gkh834},
    url = {https://doi.org/10.1093/nar/gkh834},
    eprint = {https://academic.oup.com/nar/article-pdf/32/17/5036/4156320/gkh834.pdf},
}

@Inbook{Mathews1997AnUR,
author={Mathews, David H.
and Andre, Troy C.
and Kim, James
and Turner, Douglas H.
and Zuker, Michael},
title={An Updated Recursive Algorithm for RNA Secondary Structure Prediction with Improved Thermodynamic Parameters},
series={ACS Symposium Series},
year={1997},
month={Nov},
day={26},
publisher={American Chemical Society},
volume={682},
pages={246-257},
note={0},
isbn={9780841235410},
doi={10.1021/bk-1998-0682.ch015},
url={https://doi.org/10.1021/bk-1998-0682.ch015}
}

@article{10.1073/pnas.0401799101,
author = {David H. Mathews  and Matthew D. Disney  and Jessica L. Childs  and Susan J. Schroeder  and Michael Zuker  and Douglas H. Turner },
title = {Incorporating chemical modification constraints into a dynamic programming algorithm for prediction of RNA secondary structure},
journal = {Proceedings of the National Academy of Sciences},
volume = {101},
number = {19},
pages = {7287-7292},
year = {2004},
doi = {10.1073/pnas.0401799101},
URL = {https://www.pnas.org/doi/abs/10.1073/pnas.0401799101},
eprint = {https://www.pnas.org/doi/pdf/10.1073/pnas.0401799101}}

@Article{Reuter2010,
author={Reuter, Jessica S.
and Mathews, David H.},
title={RNAstructure: software for RNA secondary structure prediction and analysis},
journal={BMC Bioinformatics},
year={2010},
month={Mar},
day={15},
volume={11},
number={1},
pages={129},
issn={1471-2105},
doi={10.1186/1471-2105-11-129},
url={https://doi.org/10.1186/1471-2105-11-129}
}

@ARTICLE{Chan2015-hs,
  title    = "{GtRNAdb} 2.0: an expanded database of transfer {RNA} genes
              identified in complete and draft genomes",
  author   = "Chan, Patricia P and Lowe, Todd M",
  journal  = "Nucleic Acids Res",
  volume   =  44,
  number   = "D1",
  pages    = "D184--9",
  month    =  dec,
  year     =  2015,
  address  = "England",
  language = "en"
}

@ARTICLE{Presnyak2015-xm,
  title    = "Codon optimality is a major determinant of {mRNA} stability",
  author   = "Presnyak, Vladimir and Alhusaini, Najwa and Chen, Ying-Hsin and
              Martin, Sophie and Morris, Nathan and Kline, Nicholas and Olson,
              Sara and Weinberg, David and Baker, Kristian E and Graveley,
              Brenton R and Coller, Jeff",
  journal  = "Cell",
  volume   =  160,
  number   =  6,
  pages    = "1111--1124",
  month    =  mar,
  year     =  2015,
  address  = "United States",
  language = "en"
}

@article {Terai2016,
	Title = {CDSfold: an algorithm for designing a protein-coding sequence with the most stable secondary structure},
	Author = {Terai, Goro and Kamegai, Satoshi and Asai, Kiyoshi},
	DOI = {10.1093/bioinformatics/btv678},
	Number = {6},
	Volume = {32},
	Month = {March},
	Year = {2016},
	Journal = {Bioinformatics (Oxford, England)},
	ISSN = {1367-4803},
	Pages = {828—834},
	URL = {https://doi.org/10.1093/bioinformatics/btv678},
}

@article{cohen2003,
author = {Cohen, Barry and Skiena, Steven},
title = {Natural Selection and Algorithmic Design of mRNA},
journal = {Journal of Computational Biology},
volume = {10},
number = {3-4},
pages = {419-432},
year = {2003},
doi = {10.1089/10665270360688101},
    note ={PMID: 12935336},

URL = { 
    
        https://doi.org/10.1089/10665270360688101
    
    

},
eprint = { 
    
        https://doi.org/10.1089/10665270360688101
    
    

}
}

@Article{Zhang2023,
author={Zhang, He
and Zhang, Liang
and Lin, Ang
and Xu, Congcong
and Li, Ziyu
and Liu, Kaibo
and Liu, Boxiang
and Ma, Xiaopin
and Zhao, Fanfan
and Jiang, Huiling
and Chen, Chunxiu
and Shen, Haifa
and Li, Hangwen
and Mathews, David H.
and Zhang, Yujian
and Huang, Liang},
title={Algorithm for optimized mRNA design improves stability and immunogenicity},
journal={Nature},
year={2023},
month={Sep},
day={01},
volume={621},
number={7978},
pages={396-403},
issn={1476-4687},
doi={10.1038/s41586-023-06127-z},
url={https://doi.org/10.1038/s41586-023-06127-z}
}

@article{li2023codonbert,
  title={Codonbert: Large language models for mrna design and optimization},
  author={Li, Sizhen and Moayedpour, Saeed and Li, Ruijiang and Bailey, Michael and Riahi, Saleh and Kogler-Anele, Lorenzo and Miladi, Milad and Miner, Jacob and Zheng, Dinghai and Wang, Jun and others},
  journal={bioRxiv},
  pages={2023--09},
  year={2023},
  publisher={Cold Spring Harbor Laboratory}
}

@misc{NCBI_Genome,
  author = {{National Center for Biotechnology Information (NCBI)}},
  title = {NCBI Genome Database},
  year = {1988--},
  url = {https://www.ncbi.nlm.nih.gov/datasets/genome/},
  note = {[cited 2025 Jan 16]}
}

@misc{dosovitskiy2021,
      title={An Image is Worth 16x16 Words: Transformers for Image Recognition at Scale}, 
      author={Alexey Dosovitskiy and Lucas Beyer and Alexander Kolesnikov and Dirk Weissenborn and Xiaohua Zhai and Thomas Unterthiner and Mostafa Dehghani and Matthias Minderer and Georg Heigold and Sylvain Gelly and Jakob Uszkoreit and Neil Houlsby},
      year={2021},
      eprint={2010.11929},
      archivePrefix={arXiv},
      primaryClass={cs.CV},
      url={https://arxiv.org/abs/2010.11929}, 
}

@article{liu2021swin,
  title={Swin transformer: Hierarchical vision transformer using shifted windows},
  author={Liu, Ze and Lin, Yutong and Cao, Yue and Hu, Han and Wei, Yixuan and Zhang, Zheng and Lin, Stephen and Guo, Baining},
  journal={arXiv preprint arXiv:2103.14030},
  year={2021}
}

@misc{loshchilov2019,
      title={Decoupled Weight Decay Regularization}, 
      author={Ilya Loshchilov and Frank Hutter},
      year={2019},
      eprint={1711.05101},
      archivePrefix={arXiv},
      primaryClass={cs.LG},
      url={https://arxiv.org/abs/1711.05101}, 
}

@article{MATHEWS1999911,
title = {Expanded sequence dependence of thermodynamic parameters improves prediction of RNA secondary structure11Edited by I. Tinoco},
journal = {Journal of Molecular Biology},
volume = {288},
number = {5},
pages = {911-940},
year = {1999},
issn = {0022-2836},
doi = {https://doi.org/10.1006/jmbi.1999.2700},
url = {https://www.sciencedirect.com/science/article/pii/S0022283699927006},
author = {David H. Mathews and Jeffrey Sabina and Michael Zuker and Douglas H. Turner}
}

@article{
david2004,
author = {David H. Mathews  and Matthew D. Disney  and Jessica L. Childs  and Susan J. Schroeder  and Michael Zuker  and Douglas H. Turner },
title = {Incorporating chemical modification constraints into a dynamic programming algorithm for prediction of RNA secondary structure},
journal = {Proceedings of the National Academy of Sciences},
volume = {101},
number = {19},
pages = {7287-7292},
year = {2004},
doi = {10.1073/pnas.0401799101},
URL = {https://www.pnas.org/doi/abs/10.1073/pnas.0401799101},
eprint = {https://www.pnas.org/doi/pdf/10.1073/pnas.0401799101},}

@misc{kingma2014adam,
      title={Adam: A Method for Stochastic Optimization}, 
      author={Diederik P. Kingma and Jimmy Ba},
      year={2017},
      eprint={1412.6980},
      archivePrefix={arXiv},
      primaryClass={cs.LG},
      url={https://arxiv.org/abs/1412.6980}, 
}

@misc{NCBI_Virus,
  author = {{National Center for Biotechnology Information (NCBI)}},
  title = {NCBI Virus Database},
  year = {1988--},
  url = {https://www.ncbi.nlm.nih.gov/labs/virus/vssi/#/},
  note = {[cited 2025 Jan 16]}
}

@incollection{KOTHARI1993119,
title = {Neural Networks for Pattern Recognition},
editor = {Marshall C. Yovits},
series = {Advances in Computers},
publisher = {Elsevier},
volume = {37},
pages = {119-166},
year = {1993},
issn = {0065-2458},
doi = {https://doi.org/10.1016/S0065-2458(08)60404-0},
url = {https://www.sciencedirect.com/science/article/pii/S0065245808604040},
author = {S.C. Kothari and Heekuck Oh},
}

@inproceedings{NEURIPS2021_cba0a4ee,
 author = {Tolstikhin, Ilya O and Houlsby, Neil and Kolesnikov, Alexander and Beyer, Lucas and Zhai, Xiaohua and Unterthiner, Thomas and Yung, Jessica and Steiner, Andreas and Keysers, Daniel and Uszkoreit, Jakob and Lucic, Mario and Dosovitskiy, Alexey},
 booktitle = {Advances in Neural Information Processing Systems},
 editor = {M. Ranzato and A. Beygelzimer and Y. Dauphin and P.S. Liang and J. Wortman Vaughan},
 pages = {24261--24272},
 publisher = {Curran Associates, Inc.},
 title = {MLP-Mixer: An all-MLP Architecture for Vision},
 volume = {34},
 year = {2021}
}

@misc{devlin2019bert,
      title={BERT: Pre-training of Deep Bidirectional Transformers for Language Understanding}, 
      author={Jacob Devlin and Ming-Wei Chang and Kenton Lee and Kristina Toutanova},
      year={2019},
      eprint={1810.04805},
      archivePrefix={arXiv},
      primaryClass={cs.CL},
      url={https://arxiv.org/abs/1810.04805}, 
}

@article{2020t5,
  author  = {Colin Raffel and Noam Shazeer and Adam Roberts and Katherine Lee and Sharan Narang and Michael Matena and Yanqi Zhou and Wei Li and Peter J. Liu},
  title   = {Exploring the Limits of Transfer Learning with a Unified Text-to-Text Transformer},
  journal = {Journal of Machine Learning Research},
  year    = {2020},
  volume  = {21},
  number  = {140},
  pages   = {1-67},
  url     = {http://jmlr.org/papers/v21/20-074.html}
}

@misc{dai2019transformerxl,
      title={Transformer-XL: Attentive Language Models Beyond a Fixed-Length Context}, 
      author={Zihang Dai and Zhilin Yang and Yiming Yang and Jaime Carbonell and Quoc V. Le and Ruslan Salakhutdinov},
      year={2019},
      eprint={1901.02860},
      archivePrefix={arXiv},
      primaryClass={cs.LG},
      url={https://arxiv.org/abs/1901.02860}, 
}

\end{document}